\documentclass[showpacs,floatfix]{revtex4}

\usepackage{graphicx,psfrag,amsmath,amssymb,amsfonts,latexsym,color,dcolumn}

\begin{document}

\title{Closed time path approach to the Casimir energy in real media}

\author{Adri\'an E. Rubio L\'opez\footnote{arubio@df.uba.ar} and Fernando C. Lombardo\footnote{lombardo@df.uba.ar}}

\affiliation{Departamento de F\'\i sica {\it Juan Jos\'e
Giambiagi}, FCEyN UBA and IFIBA CONICET-UBA, Facultad de Ciencias Exactas y Naturales,
Ciudad Universitaria, Pabell\' on I, 1428 Buenos Aires, Argentina}

\date{today}

\begin{abstract}
The closed time path (CTP) formalism is applied, in the framework of open quantum systems, to study the time evolution of the expectation value of the
energy-momentum tensor of a scalar field in the presence of real materials. We analyze quantum (Casimir) fluctuations in a fully non-equilibrium scenario, when the  scalar field is interacting with the polarization degrees of freedom of matter, described as quantum Brownian particles (harmonic oscillators coupled to a bath) in each point of space. A generalized analysis was done for two types of couplings between the field and the polarization degrees of freedom. On the one hand, we considered a bilinear coupling between the field and the polarization degrees of freedom, and on the other hand, a (more realistic) current-type coupling as in the case of the electromagnetic field interacting with matter. We successfully computed the CTP generating functional for the field, through calculating the corresponding influence functionals. We considered the high temperature limit for the field, keeping arbitrary temperatures for each part of the material's volume elements. We obtained a closed form for the Hadamard propagator, which let us study the dynamical evolution of the expectations values of the energy-momentum tensor components from the initial time, when the interactions are turned on. We showed that two contributions always take place in the transient evolution: one of these is associated to the material and the other one is only associated to the field. Transient features were studied and the long-time limit was derived in several cases. We proved that in the steady situation of a field in $n+1$ dimensions, the material always contribute unless is non-dissipative. Conversely, the proper field contribution vanishes unless the material is non-dissipative or, moreover, at least for the $1+1$ case, if there are regions without material. We finally conclude that any steady quantization scheme in $1+1$ dimensions must consider both contributions and, on the other hand, we argue why these results are physically expected from a dynamical point of view, and also could be valid for higher dimensions based on the expected continuity between the non-dissipative and real material cases
\end{abstract}


\maketitle

\newcommand{\beq}{\begin{equation}}
\newcommand{\eeq}{\end{equation}}
\newcommand{\dalam}{\nabla^2-\partial_t^2}
\newcommand{\mbf}{\mathbf}
\newcommand{\itm}{\mathit}
\newcommand{\beqa}{\begin{eqnarray}}
\newcommand{\eeqa}{\end{eqnarray}}

\section{Introduction}

The study of the Casimir forces in the framework of open quantum systems is the possibility of analyzing non-equilibrium effects, such as the Casimir force between objects at different temperatures \cite{Antezza}, the power of heat transfer between bodies \cite{Bimonte}, and the inclusion of time dependent evolutions until reaching a stationary situation. Even though, the celebrated Lifshitz formula \cite{Lifshitz} describes the forces between dielectrics at steady situation in terms of their macroscopic electromagnetic properties, this is not derived from a first principle quantum framework. The original derivation of this very general formula is based on a macroscopic approach, starting from stochastic Maxwell equations and using thermodynamical properties for the stochastic fields. As pointed out in several papers,  the connection between this approach and an approach based on a fully quantized model is not completely clear.  Moreover, some doubts have been raised about the applicability of the Lifshitz formula to lossy dielectrics \cite{Barton2010,Philbin2010,daRosaetal}.

Moreover, from a conceptual point of view, the theoretical calculations for mirrors with general electromagnetic properties, including absorption, is not a completely settled issue \cite{Barton2010,Philbin2010,daRosaetal}.  Since dissipative effects imply the possibility of energy interchanges between different parts of the full system (mirrors, vacuum field and environment), the theory of open quantum systems \cite{BreuerPett} is the natural approach to clarify the role of dissipation in Casimir physics. Indeed, in  this framework, dissipation and noise appear in the effective theory of the relevant degrees of freedom (the electromagnetic field) after integration of the matter and other environmental degrees of freedom.

The quantization at the steady situation (steady quantization scheme) can be performed starting from the macroscopic Maxwell equations,  and including noise terms to account for absorption \cite{Buhmann2007}. In this approach a canonical quantization scheme is not possible, unless one couples the electromagnetic field to a reservoir (see \cite{Philbin2010}), following the standard route to include dissipation in simple quantum mechanical systems. Another possibility is to establish a first principles model in which the slabs are described through their microscopic degrees of freedom, which are coupled to the electromagnetic field. In this kind of models,  losses are also incorporated by considering a thermal bath, to allow for  the possibility of absorption of light. There is a large body of literature on the quantization of the electromagnetic field in dielectrics. Regarding microscopic models, the fully canonical quantization of the electromagnetic field in dispersive and lossy dielectrics has been performed by Huttner and Barnett (HB) \cite{HB}. In the HB model, the electromagnetic field is coupled to matter (the polarization field), and the matter is coupled to a reservoir that is included into the model to describe the losses. In the context of the theory of quantum open systems, one can think the HB model as a composite system in which the relevant degrees of freedom belong to two subsystems (the electromagnetic field and the matter), and the matter degrees of freedom are in turn coupled to an environment (the thermal reservoir). The indirect coupling between the electromagnetic field and the thermal reservoir is responsible for the losses. As we will comment below, this will be our starting point to compute the Casimir force between absorbing media.

In a previous work \cite{LombiMazziRL}, we have followed a steady canonical quantization program similar to that of Ref.\cite{Dorota1992}, generalizing it by considering a general and well defined open quantum system. In this work, we will work with two simplified models analogous to the one of HB, both assuming that the dielectric atoms in the slabs are quantum Brownian particles, and that they are subjected to fluctuations (noise) and dissipation, due to the coupling to an external thermal environment. We will keep generality in the type of spectral density to specify the bath to which the atoms are coupled, generalizing the constant dissipation model as in Ref.\cite{LombiMazziRL}. Indeed, after integration of the environmental degrees of freedom, it will be possible to obtain the dissipation and noise kernels that modify the unitary equation of motion of the dielectric atoms.

The difference between both models relies in their couplings to the field. On the one hand, the first model, which we will call bilinear coupling model, consists in a direct coupling between the field and the atom's polarization degree of freedom in each point of space. On the other hand, the current-type coupling model consists in a coupling between the field time derivative and the atom's polarization degree of freedom. The former model is more suitable for the development of the calculations, while the latter is more realistic in the sense that is closer to the real coupling between the electromagnetic (EM) field and the matter. However, both models are of interest and can be studied in a compact way altogether to obtain general conclusions about the non-equilibrium thermodynamics and transient time evolution of quantum fields in the framework of quantum open system.

With this aim, we used the Schwinger-Keldysh formalism (or closed time path (CTP) - in-in - formalism) to provide the theoretical framework, which is based on the original papers by Schwinger \cite{Schwinger} and Keldysh \cite{Keldish} and is particularly useful for non-equilibrium quantum field theory (also see \cite{Chou, Jordan,Weinberg}). According to the CTP formalism, the expectation value of an operator and its correlation functions can be derived from an in-in generating functional in a path integral representation
\cite{CalHu}, in a similar way as it happens in the well-known in-out formalism \cite{GreiRein} but by doubling the fields and connecting them by a CTP boundary condition, which ensures that the functional derivatives of the generating functional give expectation values of the field operator. In this scheme, the open quantum systems framework is totally integrated through the concept of the influence action \cite{FeynHibbs}, resulting from partial trace over the environment degrees of freedom, giving the effective dynamics for the system through a coarse graining of the environments. Influence actions have been calculated in different context in the Literature, for example, specific models assume that during cosmological inflation the UV (or sub-Hubble) modes of a field, once integrated out, decohere the IR (or super-Hubble) modes because the former modes are inaccessible observationally. In these models, the CTP formalism applied to cosmological perturbations aims to describing the transition between the quantum nature of the initial density inhomogeneities as a consequence of inflation and the classical stochastic behaviour \cite{lombardo}.

This paper is organized as follows: In the next Section we introduce the bilinear model. In Sec. \ref{GIF}, we fully develop the CTP formalism for the open quantum system to obtain the generating functional for the field and the influence actions that result after each functional integration, identifying the dissipation and noise kernels in each influence action. In Sec. \ref{CC}, we extend, by a few number of modifications, the calculation of the generating functional done in Sec. \ref{GIF} to the current-type model, calculating the new dissipation and noise kernels. In Sec. \ref{EMTFC}, we derive a closed form for the expectation values of the energy-momentum tensor components in terms of the Hadamard propagators for each coupling model. Then, in Sec. \ref{NEBFDCOTCS} we study different scenarios of interest where our general results give different transient time behaviors and different conclusions about the steady situations in each coupling case. Finally, Sec. \ref{FR} summarize our findings. The Appendix contains some details of intermediate calculations.

\section{Bilinear Coupling}

In order to include effects of dissipation and noise (fluctuations) in the calculation of the energy density of the electromagnetic field in interaction with real media, we will develop a full CTP approach to the problem.

Therefore, we will consider a composite system consisting in two parts: the field, which we will consider a real massless scalar field and the real media, which in turn are modeled by continuous sets of quantum Brownian particles localized in certain regions of space. With this, we represent the polarization density degrees of freedom. These degrees of freedom are basically harmonic oscillators coupled to the field at each point, that can be associated with the material's atoms. The composite system (field and material atoms) is also coupled to an external bath of harmonic oscillators throught the interaction between the atoms in the material and the thermal environment.

Then, the total action for the whole system is given by

\begin{equation}
S[\phi,r,q_{n}]=S_{0}[\phi]+S_{0}[r]+\sum_{n}S_{0}[q_{n}]+S_{\rm int}[\phi,r]+\sum_{n}S_{\rm int}[r,q_{n}],
\end{equation}
where each term given by

\begin{equation}
S_{0}[\phi]=\int d\mathbf{x}\int_{t_{0}}^{t_{\rm f}}d\tau~\frac{1}{2}~\partial_{\mu}\phi~\partial^{\mu}\phi,
\label{FreeFieldAction}
\end{equation}

\begin{equation}
S_{0}[r]=\int d\mathbf{x}\int_{t_{0}}^{t_{\rm f}}d\tau~4\pi\eta_{\mathbf{x}}~g(\mathbf{x})~\frac{m_{\mathbf{x}}}{2}\left(\dot{r}_{\mathbf{x}}^{2}(\tau)-\omega_{\mathbf{x}}^{2}~r_{\mathbf{x}}^{2}(\tau)\right),
\label{FreePolAction}
\end{equation}

\begin{equation}
S_{0}[q_{n}]=\int d\mathbf{x}\int_{t_{0}}^{t_{\rm f}}d\tau~4\pi\eta_{\mathbf{x}}~g(\mathbf{x})~\frac{m_{n,\mathbf{x}}}{2}\left(\dot{q}_{n,\mathbf{x}}^{2}(\tau)-\omega_{n,\mathbf{x}}^{2}~q_{n,\mathbf{x}}^{2}(\tau)\right),
\label{FreeBathHOAction}
\end{equation}

\begin{equation}
S_{\rm int}[\phi,r]=\int d\mathbf{x}\int_{t_{0}}^{t_{\rm f}}d\tau~4\pi\eta_{\mathbf{x}}~g(\mathbf{x})~\lambda_{0,\mathbf{x}}~\phi(\mathbf{x},\tau)~r_{\mathbf{x}}(\tau),
\label{IntFieldPolAction}
\end{equation}

\begin{equation}
S_{\rm int}[r,q_{n}]=\int d\mathbf{x}\int_{t_{0}}^{t_{\rm f}}d\tau~4\pi\eta_{\mathbf{x}}~g(\mathbf{x})~\frac{\lambda_{n,\mathbf{x}}}{\sqrt{2m_{n,\mathbf{x}}~\omega_{n,\mathbf{x}}}}~r_{\mathbf{x}}(\tau)~q_{n,\mathbf{x}}(\tau),
\label{IntPolBathHOAction}
\end{equation}
where the subindex $\mathbf{x}$ denotes the fact that the oscillators (and its properties) in each point of the space are independent of each other. In other words, we have to conceive the set of oscillators $r$ associated to the polarization density that form the material more like a continuous of independent quantum degrees of freedom (with density $\eta_{\mathbf{x}}$), where each polarization degree of freedom has its own material properties (masses $m_{\mathbf{x}}$, frequency $\omega_{\mathbf{x}}$ and coupling $\lambda_{0,\mathbf{x}}$), being $\mathbf{x}$ only a label when appearing as a subindex (we are assuming that the material can be inhomogeneous). Analogously, we consider for the respective properties of each thermal bath interacting with the polarization degrees of freedom represented by the sets of oscillators $\{q_{n,\mathbf{x}}\}$ in each spatial point.

On the other hand, the matter distribution $g(\mathbf{x})$ defines the regions of material and is $g= 1$ for this regions and $g=0$ outside them.

It is also worth noting that the scalar field seems to be one of the electromagnetic field components interacting with matter. In this first model we consider, for simplicity, a bilinear coupling between the field and the polarization degree of freedom.

Finally, we will assume that the total system is initially uncorrelated, thus the initial density matrix is written as a direct product of each part, which we also suppose to be initially in a thermal equilibrium at a proper characteristic temperatures ($\beta_{\phi},\beta_{r_{\mathbf{x}}},\beta_{B,\mathbf{x}}$ -the material can also be thermally inhomogeneous-),

\begin{equation}
\widehat{\rho}(t_{0})=\widehat{\rho}_{\phi}(t_{0})\otimes\widehat{\rho}_{r_{\mathbf{x}}}(t_{0})\otimes\widehat{\rho}_{\{q_{n,\mathbf{x}}\}}(t_{0}).
\label{InitialState}
\end{equation}

\section{Generating and Influence Functionals}\label{GIF}

Our goal in this Section is to compute the  expectation value of the field quantum correlation function.  We will employ the in-in formalism by means of a closed time path (CTP) to write the field's generating functional,  after integrating out the environment by generalizing the procedure known from, for example, Refs. \cite{CalHu,CalRouVer},

\begin{eqnarray}
Z[J,J']=\int d\phi_{\rm f}\int d\phi_{0}~d\phi'_{0}\int_{\phi(\mathbf{x},t_{0})=\phi_{0}(\mathbf{x})}^{\phi(\mathbf{x},t_{\rm f})=\phi_{\rm f}(\mathbf{x})}\mathcal{D}\phi\int_{\phi'(\mathbf{x},t_{0})=\phi'_{0}(\mathbf{x})}^{\phi'(\mathbf{x},t_{\rm f})=\phi_{\rm f}(\mathbf{x})}\mathcal{D}\phi'&&\rho_{\phi}(\phi_{0},\phi'_{0},t_{0})~e^{i\left(S_{0}[\phi]-S_{0}[\phi']\right)}~\mathcal{F}[\phi,\phi']\nonumber\\
&&\times~ e^{i\int d\mathbf{x}\int_{t_{0}}^{t_{\rm f}}d\tau\left(J(\mathbf{x},\tau)~\phi(\mathbf{x},\tau)-J'(\mathbf{x},\tau)~\phi'(\mathbf{x},\tau)\right)},
\label{GeneratingFunctional}
\end{eqnarray}
where the field's functional $\mathcal{F}$ is known as the influence functional \cite{FeynHibbs} which is related to the field's influence action $S_{\rm IF}[\phi,\phi']$ generated by the material degrees of freedom (atoms plus baths).

Since the material is modeled as a continuum of spatially-independent oscillators each one interacting with its own bath, the influence functional clearly factorizes in the spatial label resulting

\begin{eqnarray}
\mathcal{F}[\phi,\phi']=e^{iS_{\rm IF}[\phi,\phi']}&=&\prod_{\mathbf{x}}\int dr_{\rm f,\mathbf{x}}\int dr_{0,\mathbf{x}}~dr'_{0,\mathbf{x}}\int_{r_{\mathbf{x}}(t_{0})=r_{0,\mathbf{x}}}^{r_{\mathbf{x}}(t_{\rm f})=r_{\rm f,\mathbf{x}}}\mathcal{D}r_{\mathbf{x}}\int_{r'_{\mathbf{x}}(t_{0})=r'_{0,\mathbf{x}}}^{r'_{\mathbf{x}}(t_{\rm f})=r_{\rm f,\mathbf{x}}}\mathcal{D}r'_{\mathbf{x}} ~ \rho_{r_{\mathbf{x}}}\left(r_{0,\mathbf{x}},r'_{0,\mathbf{x}},t_{0}\right)~\nonumber\\
&\times & e^{i4\pi\eta_{\mathbf{x}}g(\mathbf{x})\left(S_{0}[r_{\mathbf{x}}]-S_{0}[r'_{\mathbf{x}}]\right)}~e^{i4\pi\eta_{\mathbf{x}}g(\mathbf{x})~S_{\rm QBM}[r_{\mathbf{x}},r'_{\mathbf{x}}]}~e^{i4\pi\eta_{\mathbf{x}}g(\mathbf{x})\left(S_{\rm int}[\phi,r_{\mathbf{x}}]-S_{\rm int}[\phi',r'_{\mathbf{x}}]\right)}.
\label{InfluenceFuntionalField}
\end{eqnarray}
where $S_{\rm QBM}[r_{\mathbf{x}},r'_{\mathbf{x}}]=\int_{t_{0}}^{t_{\rm f}}d\tau\int_{t_{0}}^{t_{\rm f}}d\tau'~\Delta r_{\mathbf{x}}(\tau)\left(-2~D_{\rm QBM,\mathbf{x}}(\tau-\tau')~\Sigma r_{\mathbf{x}}(\tau')+\frac{i}{2}~N_{\rm QBM,\mathbf{x}}(\tau-\tau')~\Delta r_{\mathbf{x}}(\tau')\right)$ is the well-known influence action for the QBM theory \cite{CaldeLegg,HuPazZhang}, which represents the influence of a bath at $\mathbf{x}$ (given by the set $\{q_{n,\mathbf{x}}\}$) over the polarization degrees of freedom $r_{\mathbf{x}}$ at the same spatial point. It is worth noting that in this expression the scalar fields $\phi$ and $\phi'$ appear as additional external sources as $J$ and $J'$ were so for the field. It is worth noting that we have set $\Delta r_{\mathbf{x}}=r'_{\mathbf{x}}-r_{\mathbf{x}}$ and $\Sigma r_{\mathbf{x}}=\left(r_{\mathbf{x}}+r'_{\mathbf{x}}\right)/2$, and that the QBM's influence action is clearly the analogous result of the CTP expression for the influence functional of the field of Eq.(\ref{InfluenceFuntionalField}), where the trace has been taken over the bath's degree of freedom $\{q_{n,\mathbf{x}}\}$ considering them in a thermal state.

The kernels $N_{\rm QBM,\mathbf{x}}$ and $D_{\rm QBM,\mathbf{x}}$ in $S_{\rm QBM}$ are nothing more than the QBM noise and dissipation kernels respectively \cite{CalHu,HuPazZhang}. It is clear that the expression of the influence action is quite general and applies to all type of baths (characterized by the spectral density being subohmic, ohmic or supraohmic \cite{HuPazZhang,BreuerPett}) characterized by a particular temperature. In the same way, it turns out that the noise kernel $N_{\rm QBM,\mathbf{x}}$ corresponds to the sum of the Hadamard propagators for the bath oscillators at the point $\mathbf{x}$, while the dissipation kernel $D_{\rm QBM,\mathbf{x}}$ corresponds to the sum of the retarded propagators at the same point, which clearly shows a causal behavior ($D_{\rm QBM,\mathbf{x}}(\tau,\tau')\propto\Theta(\tau-\tau')$).

\subsection{Field's Influence Functional}

At this point, we have to compute the influence functional for the field, $\mathcal{F}$ of Eq.(\ref{InfluenceFuntionalField}). For this purpose, we have to evaluate each factor in the product. The result of this type of CTP integrals can be found in Ref. \cite{CalRouVer}. We present a generalization due to consider the degrees of freedom of the polarization density which is straightforward (polarization or bath degrees of freedom must contain a dimensional normalization factor $\frac{1}{4\pi\eta_{\mathbf{x}}}$ -see Ref. \cite{LombiMazziRL}- and we also have to take in account that the matter distribution satisfies $g^{2}(\mathbf{x})=g(\mathbf{x})$), therefore we obtain,

\begin{eqnarray}
\mathcal{F}[\phi,\phi']&=&\prod_{\mathbf{x}}\Big\langle e^{-i4\pi\eta_{\mathbf{x}}g(\mathbf{x})~\lambda_{0,\mathbf{x}}\int_{t_{0}}^{t_{\rm f}}d\tau~\Delta\phi(\mathbf{x},\tau)~\mathcal{R}_{0,\mathbf{x}}(\tau)}\Big\rangle_{r_{0,\mathbf{x}},p_{0,\mathbf{x}}}e^{-\frac{1}{2}~4\pi\eta_{\mathbf{x}}g(\mathbf{x})\int_{t_{0}}^{t_{\rm f}}d\tau\int_{t_{0}}^{t_{\rm f}}d\tau'~\Delta\phi(\mathbf{x},\tau)~\mathcal{N}_{B,\mathbf{x}}(\tau,\tau')~\Delta\phi(\mathbf{x},\tau')}\nonumber\\
&&\times~e^{-i4\pi\eta_{\mathbf{x}}g(\mathbf{x})\int_{t_{0}}^{t_{\rm f}}d\tau\int_{t_{0}}^{t_{\rm f}}d\tau'~\Delta\phi(\mathbf{x},\tau)~2~\mathcal{D}_{\mathbf{x}}(\tau,\tau')~\Sigma\phi(\mathbf{x},\tau')},
\label{CTPIntPol}
\end{eqnarray}
where $\Delta\phi=\phi'-\phi$, $\Sigma\phi=(\phi+\phi')/2$ and  $\mathcal{D}_{\mathbf{x}}(\tau,\tau')\equiv\mathcal{D}_{\mathbf{x}}(\tau-\tau')=\frac{\lambda_{0,\mathbf{x}}^{2}}{2}~G_{\rm Ret,\mathbf{x}}(\tau-\tau')$ is the dissipation kernel over the field, being $G_{\rm Ret,\mathbf{x}}$ the retarded Green function and $\mathcal{R}_{0,\mathbf{x}}$ the solution with initial conditions $\{r_{0,\mathbf{x}},p_{0,\mathbf{x}}\}$ associated to the semiclassical equation of motion, that results from the homogeneous equation

\begin{eqnarray}
\frac{\delta S_{\rm CTP}[r_{\mathbf{x}},r'_{\mathbf{x}}]}{\delta r_{\mathbf{x}}}\Big|_{r_{\mathbf{x}}=r'_{\mathbf{x}}}=\frac{\delta S_{\rm CTP}[\Delta r_{\mathbf{x}},\Sigma r_{\mathbf{x}}]}{\delta \Delta r_{\mathbf{x}}}\Big|_{\Delta r_{\mathbf{x}}=0}&=&0,\nonumber\\
\ddot{r}_{\mathbf{x}}+\omega_{\mathbf{x}}^{2}~r_{\mathbf{x}}-\frac{2}{m_{\mathbf{x}}}\int_{t_{0}}^{t}d\tau~D_{\rm QBM,\mathbf{x}}(t-\tau)~r_{\mathbf{x}}(\tau)&=&0,
\label{EqMotionR}
\end{eqnarray}
where $S_{\rm CTP}[r_{\mathbf{x}},r'_{\mathbf{x}}]=S_{0}[r_{\mathbf{x}}]-S_{0}[r'_{\mathbf{x}}]+S_{\rm QBM}[r_{\mathbf{x}},r'_{\mathbf{x}}]$ and,

\begin{equation}
\mathcal{R}_{0,\mathbf{x}}(\tau)=r_{0,\mathbf{x}}~\dot{G}_{\rm Ret,\mathbf{x}}(\tau-t_{0})+\frac{p_{0,\mathbf{x}}}{m_{\mathbf{x}}}~G_{\rm Ret,\mathbf{x}}(\tau-t_{0}).
\label{HomoSolQBM}
\end{equation}

On the other hand, the kernel $\mathcal{N}_{B,\mathbf{x}}$ is the part of the noise kernel associated to the baths that acts on the field (there is another part associated to the first factor in the right hand side of Eq.(\ref{CTPIntPol})),

\begin{equation}
\mathcal{N}_{B,\mathbf{x}}(\tau,\tau')=\lambda_{0,\mathbf{x}}^{2}\int_{t_{0}}^{t_{\rm f}}ds\int_{t_{0}}^{t_{\rm f}}ds'~G_{\rm Ret,\mathbf{x}}(\tau-s)~N_{\rm QBM,\mathbf{x}}\left(s-s'\right)~G_{\rm Ret,\mathbf{x}}(\tau'-s').
\label{PhiNoiseKernelB}
\end{equation}

Finally, the first factor in the right hand side of Eq.(\ref{CTPIntPol}) is given by (see Ref. \cite{CalRouVer})

\begin{eqnarray}
\Big\langle e^{-i4\pi\eta_{\mathbf{x}}g(\mathbf{x})~\lambda_{0,\mathbf{x}}\int_{t_{0}}^{t_{\rm f}}d\tau~\Delta\phi(\mathbf{x},\tau)~\mathcal{R}_{0,\mathbf{x}}(\tau)}\Big\rangle_{r_{0,\mathbf{x}},p_{0,\mathbf{x}}} &=& \int dr_{0,\mathbf{x}}\int dp_{0,\mathbf{x}} e^{-i4\pi\eta_{\mathbf{x}}g(\mathbf{x})~\lambda_{0,\mathbf{x}}\int_{t_{0}}^{t_{\rm f}}d\tau~\Delta\phi(\mathbf{x},\tau)~\mathcal{R}_{0,\mathbf{x}}(\tau)} \nonumber \\
&\times& W_{r_{\mathbf{x}}}\left(r_{0,\mathbf{x}},p_{0,\mathbf{x}},t_{0}\right),\label{eq14}
\end{eqnarray}
where $W_{r_{\mathbf{x}}}\left(r_{0,\mathbf{x}},p_{0,\mathbf{x}},t_{0}\right)$ is the Wigner functional associated to the density matrix of the polarization degrees of freedom $\widehat{\rho}_{r_{\mathbf{x}}}(t_{0})$. This functional can be written by generalizing the expression found in Ref. \cite{CalRouVer},

\begin{equation}
W_{r_{\mathbf{x}}}\left(r_{0,\mathbf{x}},p_{0,\mathbf{x}},t_{0}\right)=\frac{1}{2\pi}\int_{-\infty}^{+\infty}d\Gamma~e^{i4\pi\eta_{\mathbf{x}}g(\mathbf{x})~p_{0,\mathbf{x}}\Gamma}~\rho_{r_{\mathbf{x}}}\left(r_{0,\mathbf{x}}-\frac{\Gamma}{2},r_{0,\mathbf{x}}+\frac{\Gamma}{2},t_{0}\right).
\end{equation}

Considering thermal initial states for each part of the total composite system, we take the density matrices for the polarization degrees of freedom to be Gaussian functions. Therefore, Eq.(\ref{eq14}) also Gaussian since the Wigner function is Gaussian in $r_{0,\mathbf{x}}$ and $p_{0,\mathbf{x}}$. This way, by considering Eq.(\ref{HomoSolQBM}), we can easily calculate the first factor on the right hand of Eq.(\ref{CTPIntPol}) as

\begin{eqnarray}
\Big\langle e^{-i4\pi\eta_{\mathbf{x}}g(\mathbf{x})~\lambda_{0,\mathbf{x}}\int_{t_{0}}^{t_{\rm f}}d\tau~\Delta\phi(\mathbf{x},\tau)~\mathcal{R}_{0,\mathbf{x}}(\tau)}\Big\rangle_{r_{0,\mathbf{x}},p_{0,\mathbf{x}}}  &=& \frac{1}{4\pi\eta_{\mathbf{x}}g(\mathbf{x})~2\sinh\left(\frac{\beta_{r_{\mathbf{x}}}\omega_{\mathbf{x}}}{2}\right)} \nonumber \\
&\times& e^{-\frac{1}{2}~4\pi\eta_{\mathbf{x}}g(\mathbf{x})\int_{t_{0}}^{t_{\rm f}}d\tau\int_{t_{0}}^{t_{\rm f}}d\tau'~\Delta\phi(\mathbf{x},\tau)~\mathcal{N}_{r,\mathbf{x}}(\tau,\tau')~\Delta\phi(\mathbf{x},\tau')},
\label{FirstFactorPol}
\end{eqnarray}
with

\begin{eqnarray}
\mathcal{N}_{r,\mathbf{x}}(\tau,\tau')=\frac{\lambda_{0,\mathbf{x}}^{2}}{2m_{\mathbf{x}}\omega_{\mathbf{x}}}~\coth\left(\frac{\beta_{r_{\mathbf{x}}}\omega_{\mathbf{x}}}{2}\right)\left[\dot{G}_{\rm Ret,\mathbf{x}}(\tau-t_{0})~\dot{G}_{\rm Ret,\mathbf{x}}(\tau'-t_{0})+\omega_{\mathbf{x}}^{2}~G_{\rm Ret,\mathbf{x}}(\tau-t_{0})~G_{\rm Ret,\mathbf{x}}(\tau'-t_{0})\right],
\label{PhiNoiseKernelR}
\end{eqnarray}
which is the other part of the noise kernel that acts on the field.  This is associated to the influence generated by the polarization degrees of freedom (it carries a global thermal factor containing the temperature of the polarization degrees of freedom $\beta_{r_{\mathbf{x}}}$).

Hence, after the normalization procedure of $Z[J,J']$, Eq.(\ref{CTPIntPol}) finally reads

\begin{equation}
\mathcal{F}[\phi,\phi']=e^{iS_{\rm IF}[\phi,\phi']},
\end{equation}
with

\begin{eqnarray}
S_{\rm IF}[\phi,\phi']&=&\int d\mathbf{x}\int_{t_{0}}^{t_{\rm f}}d\tau\int_{t_{0}}^{t_{\rm f}}d\tau'~4\pi\eta_{\mathbf{x}}~g(\mathbf{x})~\Delta\phi(\mathbf{x},\tau)\left[-2~\mathcal{D}_{\mathbf{x}}(\tau-\tau')~\Sigma\phi(\mathbf{x},\tau')+\frac{i}{2}~\mathcal{N}_{\mathbf{x}}(\tau,\tau')~\Delta\phi(\mathbf{x},\tau')\right]\nonumber\\
&=&\int d^{4}x\int d^{4}x'~\Delta\phi(x)\left[-2~\mathcal{D}(x,x')~\Sigma\phi(x')+\frac{i}{2}~\mathcal{N}(x,x')~\Delta\phi(x')\right],
\label{FieldSIF}
\end{eqnarray}
where in the last line $\mathcal{D}(x,x')\equiv 4\pi\eta_{\mathbf{x}}g(\mathbf{x})~\delta(\mathbf{x}-\mathbf{x}')~\mathcal{D}_{\mathbf{x}}(\tau-\tau')$ and $\mathcal{N}(x,x')\equiv 4\pi\eta_{\mathbf{x}}g(\mathbf{x})~\delta(\mathbf{x}-\mathbf{x}')~\mathcal{N}_{\mathbf{x}}(\tau,\tau')$ (with $\mathcal{N}_{\mathbf{x}}(\tau,\tau')=\mathcal{N}_{r,\mathbf{x}}(\tau,\tau')+\mathcal{N}_{B,\mathbf{x}}(\tau,\tau')$) for the dissipation and noise kernels respectively. The four-dimensional translational symmetry is broken by the spatial coordinates, because the $n+1$ field is interacting with $0+1$ fields, the polarization degrees of freedom. This causes that the temporal and spatial coordinates are not in equal footing.

As expected for linear couplings, the influence action for the field have the same form as $S_{QBM}$ obtained after bath's integration but for
a field in four dimensions all over the space (this is not only true for bilinear couplings between the coordinates, it is also true for bilinear couplings between a coordinate and a momenta, but logically the kernels change as we will see in next sections).

\subsection{CTP Generating Functional}

We have achieved an exact result for the influence functional and, consequently, for the influence action. Thus, going back to Eq.(\ref{GeneratingFunctional}) we can note that this CTP integrals are of the form of  Eq.(\ref{CTPIntPol}), replacing the degree of freedom by a scalar field.
Generalizing the result found in Ref. \cite{CalRouVer} for fields, we get

\begin{eqnarray}
Z[J,J']&=&\Big\langle e^{-i\int d^{4}x~J_{\Delta}(x)~\Phi_{0}(x)}\Big\rangle_{\phi_{0}(\mathbf{x}),\Pi_{0}(\mathbf{x})}~e^{-\frac{1}{2}\int d^{4}x\int d^{4}x'\int d^{4}y'\int d^{4}y~J_{\Delta}(x)~\mathcal{G}_{\rm Ret}(x,x')~\mathcal{N}(x',y')~\mathcal{G}_{\rm Ret}(y,y')~J_{\Delta}(y)}\nonumber\\
&\times & ~e^{-i\int d^{4}x\int d^{4}y~J_{\Delta}(x)~\mathcal{G}_{\rm Ret}(x,y)~J_{\Sigma}(y)},
\label{GeneratingFunctionalNOFinal}
\end{eqnarray}
where $\phi_{0}(\mathbf{x})=\phi(\mathbf{x},t_{0})$ and $\Pi_{0}(\mathbf{x})=\dot{\phi}(\mathbf{x},t_{0})$ are the initial conditions for the field, while $J_{\Delta}=J'-J$ and $J_{\Sigma}=(J+J')/2$.

Analogously to the integration performed in the last section, $\mathcal{G}_{\rm Ret}$ is a retarded Green function, this time associated to the field's semiclassical equation that results from the homogeneous equation of motion for the CTP effective action for the field: $S_{\rm CTP}[\phi,\phi']=S_{0}[\phi]-S_{0}[\phi']+S_{IF}[\phi,\phi']$,

\begin{eqnarray}
\frac{\delta S_{\rm CTP}[\phi,\phi']}{\delta\phi}\Big|_{\phi=\phi'}=\frac{\delta S_{\rm CTP}[\Delta\phi,\Sigma \phi]}{\delta \Delta\phi}\Big|_{\Delta\phi=0}&=&0,\nonumber\\
\partial_{\mu}\partial^{\mu}\phi-2\int d^{4}x'~\mathcal{D}(x,x')~\phi(x')&=&0.
\label{EqMotionPhi}
\end{eqnarray}

In the same way, $\Phi_{0}(x)$ is the solution of last equation that satisfies the initial condition $\{\phi_{0}(\mathbf{x}),\Pi_{0}(\mathbf{x})\}$, i. e.:

\begin{equation}
\Phi_{0}(x)=\int d\mathbf{x}'~\dot{\mathcal{G}}_{\rm Ret}(\mathbf{x},\mathbf{x}',t-t_{0})~\phi_{0}(\mathbf{x}')+\int d\mathbf{x}'~\mathcal{G}_{\rm Ret}(\mathbf{x},\mathbf{x}',t-t_{0})~\Pi_{0}(\mathbf{x}').
\end{equation}

To calculate the first factor involving the average over the initial conditions, we use

\begin{eqnarray}
\Big\langle e^{-i\int d^{4}x~J_{\Delta}(x)~\Phi_{0}(x)}\Big\rangle_{\phi_{0}(\mathbf{x}),\Pi_{0}(\mathbf{x})}&=&\int \mathcal{D}\phi_{0}(\mathbf{x}')\int \mathcal{D}\Pi_{0}(\mathbf{x}')~W_{\phi}\left[\phi_{0}(\mathbf{x}'),\Pi_{0}(\mathbf{x}'),t_{0}\right]\nonumber\\
&&\times~e^{-i\int d\mathbf{x}'\int d^{4}x~J_{\Delta}(x)\left[\dot{\mathcal{G}}_{\rm Ret}(\mathbf{x},\mathbf{x}',\tau-t_{0})~\phi_{0}(\mathbf{x}')+\mathcal{G}_{\rm Ret}(\mathbf{x},\mathbf{x}',\tau-t_{0})~\Pi_{0}(\mathbf{x}')\right]},
\label{InitialCalzettaFactor}
\end{eqnarray}
where $W_{\phi}\left[\phi_{0}(\mathbf{x}'),\Pi_{0}(\mathbf{x}'),t_{0}\right]$ plays the same role that the Wigner function in Ref.\cite{MrowMull}.

\subsubsection{Initial State Contribution of the field}

Once we have calculated the Wigner functional for the field in a thermal state (Eq.(\ref{FieldWignerCoordinate})), we go back to Eq.(\ref{InitialCalzettaFactor}) to finally calculate the first factor in the right hand of Eq.(\ref{GeneratingFunctionalNOFinal}).

For an arbitrary value for the field's temperature, the factor, which in principle are functional integrals over the field $\phi_{0}(\mathbf{x})$ and its associated momentum $\Pi_{0}(\mathbf{x})$, splits in each functional integration because the exponent also separated in each variables, therefore

\begin{eqnarray}
\Big\langle e^{-i\int d^{4}x~J_{\Delta}(x)~\Phi_{0}(x)}\Big\rangle_{\phi_{0}(\mathbf{x}),\Pi_{0}(\mathbf{x})}&=&\int\mathcal{D}\phi_{0}(\mathbf{x})~e^{-\frac{\beta_{\phi}}{2}\int d\mathbf{x}\int d\mathbf{x}'~\Delta_{\beta_{\phi}}(\mathbf{x}-\mathbf{x}')~\nabla\phi_{0}(\mathbf{x})\cdot\nabla\phi_{0}(\mathbf{x}')}~e^{\beta_{\phi}\int d\mathbf{x}~\mathcal{J}_{\phi}(\mathbf{x})~\phi_{0}(\mathbf{x})}\nonumber\\
&&\times\int\mathcal{D}\Pi_{0}(\mathbf{x})~e^{-\frac{\beta_{\phi}}{2}\int d\mathbf{x}\int d\mathbf{x}'~\Delta_{\beta_{\phi}}(\mathbf{x}-\mathbf{x}')~\Pi_{0}(\mathbf{x})~\Pi_{0}(\mathbf{x}')}~e^{\beta_{\phi}\int d\mathbf{x}~\mathcal{J}_{\Pi}(\mathbf{x})~\Pi_{0}(\mathbf{x})},
\label{InitialFactorFunctional}
\end{eqnarray}
where:

\begin{eqnarray}
\mathcal{J}_{\phi}(\mathbf{x})\equiv-\frac{i}{\beta_{\phi}}\int d^{4}x'~J_{\Delta}(x')~\dot{\mathcal{G}}_{\rm Ret}\left(\mathbf{x}',\mathbf{x},t'-t_{0}\right),
\end{eqnarray}

\begin{eqnarray}
\mathcal{J}_{\Pi}(\mathbf{x})\equiv-\frac{i}{\beta_{\phi}}\int d^{4}x'~J_{\Delta}(x')~\mathcal{G}_{\rm Ret}\left(\mathbf{x}',\mathbf{x},t'-t_{0}\right).
\end{eqnarray}

Both functional integrals will define the contribution of the first factor to the generating functional of Eq.(\ref{FieldWignerCoordinate}). In fact, it will define the contribution of the initial state of the field to the dynamical evolution, relaxation and steady situation of the system.

\subsubsection{High Temperature Limit}

First of all, to continue the calculation, we can explore the high temperature limit for the field, which seems to be the easier case to solve the functional integrals in Eq.(\ref{InitialFactorFunctional}). The high temperature approximation is given by  $\frac{\beta_{\phi}|\mathbf{p}|}{2}<<1$ on the thermal weight in momentum space (Eq.(\ref{ThermalWeightMomentum})). Then, $\tanh\left(\frac{\beta_{\phi}|\mathbf{p}|}{2}\right)\approx\frac{\beta_{\phi}|\mathbf{p}|}{2}$, and the thermal weight in momentum space is approximately $1$. Thus, in the coordinate space:

\begin{eqnarray}
\Delta_{\beta_{\phi}}(\mathbf{x}'-\mathbf{x}'')\approx\int\frac{d\mathbf{p}}{(2\pi)^{3}}~e^{-i\mathbf{p}\cdot\left(\mathbf{x}'-\mathbf{x}''\right)}\equiv\delta\left(\mathbf{x}'-\mathbf{x}''\right).
\end{eqnarray}

In this approximation, Eq.(\ref{InitialFactorFunctional}) simplifies because one integral in the exponents is straightforwardly evaluated. In this limit, both functional integrals are easily calculated, and in fact the integration over the momentum $\Pi_{0}(\mathbf{x})$, is simply a Gaussian

\begin{eqnarray}
&&\int\mathcal{D}\Pi_{0}(\mathbf{x})~e^{-\frac{\beta_{\phi}}{2}\int d\mathbf{x}~\Pi_{0}(\mathbf{x})~\Pi_{0}(\mathbf{x})}~e^{\beta_{\phi}\int d\mathbf{x}~\mathcal{J}_{\Pi}(\mathbf{x})~\Pi_{0}(\mathbf{x})}\nonumber\\
&=&e^{-\frac{1}{2\beta_{\phi}}\int d^{4}y\int d^{4}y'~J_{\Delta}(y)\left[\int d\mathbf{x}~\mathcal{G}_{\rm Ret}\left(\mathbf{y},\mathbf{x},\tau-t_{0}\right)~\mathcal{G}_{\rm Ret}\left(\mathbf{y}',\mathbf{x},\tau'-t_{0}\right)\right]J_{\Delta}(y')},
\end{eqnarray}
where in this notation $y=(\tau,\mathbf{y}),y'=(\tau',\mathbf{y}')$ and we are discarding any normalization constant that will eventually go away, at the end, in the normalization of the generating functional.

At this point, it is interesting that the high temperature approximation seems to erase all the differences between the result obtained for a single problem due to the number of spatial dimensions as it was remarked at the end of the last section. This can be noted in the fact that the thermal weight, in this limit, turns out to be the Dirac delta in all the coordinates in question, independently of the spatial dimensionality, and thus all the possible differences due to the different functional forms of the thermal weight on a given number of dimensions, seems to disappear. However, the dimensionality appears again to make differences when the functional integral over $\phi_{0}(\mathbf{x})$ has to be solved. That integral is a Gaussian functional integral too. Then, we can proceed by integrating by parts the exponent involving gradients and discard terms involving the vanishing asymptotic decay of the field at infinity ($\phi_{0}(x_{i}=\pm\infty)\rightarrow 0$). Therefore, the functional integral over $\phi_{0}(\mathbf{x})$, in the high temperature limit, is a simple Gaussian functional integral, finally obtaining

\begin{eqnarray}
\int\mathcal{D}\phi_{0}(\mathbf{x})~e^{-\frac{\beta_{\phi}}{2}\int d\mathbf{x}~\nabla\phi_{0}(\mathbf{x})\cdot\nabla\phi_{0}(\mathbf{x})}&&e^{\beta_{\phi}\int d\mathbf{x}~\mathcal{J}_{\phi}(\mathbf{x})~\phi_{0}(\mathbf{x})}\propto e^{\frac{1}{2}\int d\mathbf{x}\int d\mathbf{x}'~\beta_{\phi}^{2}~\mathcal{J}_{\phi}(\mathbf{x}')~K\left(\mathbf{x},\mathbf{x}'\right)~\mathcal{J}_{\phi}(\mathbf{x}')}\nonumber\\
&=&e^{-\frac{1}{2}\int d^{4}y\int d^{4}y'~J_{\Delta}(y)\left[\int d\mathbf{x}\int d\mathbf{x}'~\dot{\mathcal{G}}_{\rm Ret}(\mathbf{y},\mathbf{x},\tau-t_{0})~K(\mathbf{x}-\mathbf{x}')~\dot{\mathcal{G}}_{\rm Ret}(\mathbf{y}',\mathbf{x}',\tau'-t_{0})\right]J_{\Delta}(y')},
\end{eqnarray}
where $K(\mathbf{x},\mathbf{x}')=\left(-\beta_{\phi}\nabla^{2}\right)^{-1}$ is the inverse of the Laplace operator, i. e., is the Green function defined by

\begin{equation}
-\beta_{\phi}~\nabla^{2}K(\mathbf{x},\mathbf{x}')=\delta(\mathbf{x}-\mathbf{x}').
\end{equation}
It is clear the kernel has to depend on $\mathbf{x}-\mathbf{x}'$.

Since the equation is analogous to the one for the Green function of a point charge in free space (although the thermal factor appears to be as a constant permittivity), we can solve the equation taking the Fourier transform

\begin{equation}
K(\mathbf{x}-\mathbf{x}')=\int\frac{d\mathbf{p}}{(2\pi)^{3}}~e^{-i\mathbf{p}\cdot\left(\mathbf{x}-\mathbf{x}'\right)}~\overline{K}(\mathbf{p}),
\label{KernelKHighTCoordinate}
\end{equation}
where

\begin{equation}
\overline{K}(\mathbf{p})=\frac{1}{\beta_{\phi}~|\mathbf{p}|^{2}}.
\label{KernelKHighTMomentum}
\end{equation}

It is worth noting that the kernel $K(\mathbf{x}-\mathbf{x}')$ strongly depends on the dimensionality of the problem, so the number of dimensions in the problem could modify the final results.

Finally, the first factor on the generating functional in the high temperature limit results

\begin{eqnarray}
\Big\langle e^{-i\int d^{4}x~J_{\Delta}(x)~\Phi_{0}(x)}\Big\rangle_{\phi_{0}(\mathbf{x}),\Pi_{0}(\mathbf{x})}&=&e^{-\frac{1}{2}\int d^{4}y\int d^{4}y'~J_{\Delta}(y)\left[\mathcal{A}(y,y')+\mathcal{B}(y,y')\right]J_{\Delta}(y')},
\label{InitialFactorFunctionalHighTFINAL}
\end{eqnarray}
where the kernels are:

\begin{equation}
\mathcal{A}(y,y')\equiv\frac{1}{\beta_{\phi}}\int d\mathbf{x}~\mathcal{G}_{\rm Ret}\left(\mathbf{y},\mathbf{x},\tau-t_{0}\right)~\mathcal{G}_{\rm Ret}\left(\mathbf{y}',\mathbf{x},\tau'-t_{0}\right),
\label{KernelAHighT}
\end{equation}

\begin{equation}
\mathcal{B}(y,y')\equiv\int d\mathbf{x}\int d\mathbf{x}'~\dot{\mathcal{G}}_{\rm Ret}(\mathbf{y},\mathbf{x},\tau-t_{0})~K(\mathbf{x}-\mathbf{x}')~\dot{\mathcal{G}}_{\rm Ret}(\mathbf{y}',\mathbf{x}',\tau'-t_{0}).
\label{KernelBHighT}
\end{equation}

The result is symmetric, i. e., $\mathcal{A}(y,y')=\mathcal{A}(y',y)$ and $\mathcal{B}(y,y')=\mathcal{B}(y',y)$ and  we can clearly note that both kernels depend linearly with the field initial temperature as we expected from the high temperature approximation. The presence of kernels $\mathcal{A}(y,y')$ and $\mathcal{B}(y,y')$ is one of the main results of this article. We will remark their role in the Casimir energy density, and in the
contribution to the energy in the long time regime.

All in all, we now can finally write the normalized generating functional for the field in the high temperature limit, by inserting Eq.(\ref{InitialFactorFunctionalHighTFINAL}) in Eq.(\ref{GeneratingFunctionalNOFinal}),

\begin{eqnarray}
Z[J,J']&=&e^{-\frac{1}{2}\int d^{4}y\int d^{4}y'~J_{\Delta}(y)\left[\mathcal{A}(y,y')+\mathcal{B}(y,y')+\int d^{4}x\int d^{4}x'~\mathcal{G}_{\rm Ret}(y,x)~\mathcal{N}(x,x')~\mathcal{G}_{\rm Ret}(y',x')\right]J_{\Delta}(y')}\nonumber\\
&&\times~e^{-i\int d^{4}y\int d^{4}y'~J_{\Delta}(y)~\mathcal{G}_{\rm Ret}(y,y')~J_{\Sigma}(y')},
\label{GeneratingFunctionalHighTFINAL}
\end{eqnarray}
where it is worth noting that the first factor on the right hand side is accompanied by two $J_{\Delta}$, whereas that the second one is accompanied by one $J_{\Delta}$ and one $J_{\Sigma}$. This difference will make that the first and third exponents will contribute to the energy while the second one will not.

Finally, we have calculated the field generating functional in a fully dynamical scenario in the high temperature limit for the field.  This was done keeping the polarization degrees of freedom volume elements and its baths, with their own properties and temperatures. However, the model contains a bilinear interaction between the matter and the field. In the next section, we will see how to obtain, straightforwardly, the generating functional for the case of a more realistic model, i. e., a current-type interaction.

\section{Current-type Coupling}\label{CC}

At this point, we have calculated the generating functional for a massless scalar field interacting with matter as Brownian particles. It is clear that in the calculation done in the previous Sections, the field and the polarization degrees of freedom are coupled linearly, i. e., the coupling is directly on the quantum degrees of freedom. Therefore, that model is not a scalar version for one of the electromagnetic field components interacting with matter, since the interaction is not a current-type interaction. Therefore, in this Section, we will show how to extend the calculation to the case of a current-type interaction between the matter and the field, getting closer to a realistic electromagnetic model.

Then, we have to start by replacing the interaction action $S_{\rm int}[\phi,r]$ between the field and the matter in Eq.(\ref{IntFieldPolAction}) by a current-type interaction term as

\begin{equation}
\widetilde{S}_{\rm int}[\phi,r]=\int d\mathbf{x}\int_{t_{0}}^{t_{\rm f}}d\tau~4\pi\eta_{\mathbf{x}}~g(\mathbf{x})~\lambda_{0,\mathbf{x}}~\dot{\phi}(\mathbf{x},\tau)~r_{\mathbf{x}}(\tau)\equiv S_{\rm int}[\dot{\phi},r],
\label{InteractionFieldPolCurrent}
\end{equation}
where $\lambda_{0,\mathbf{x}}$ effectively plays the role of the electric charge in the electromagnetic model. It is also worth noting that we write the time derivative acting on the field, instead on the polarization degree of freedom. Both choices lead to the same equations of motion for the composite system so they are physically equivalent.  In fact, all the calculations of the last Section, devoted to calculate the field influence action, are formally the same, and we can obtain it, in principle, by simply replacing $\phi$ by $\dot{\phi}$. Therefore, the influence action on the field, in this case, reads

\begin{eqnarray}
\widetilde{S}_{\rm IF}[\phi,\phi']\equiv S_{\rm IF}[\dot{\phi},\dot{\phi}']&=&\int d\mathbf{x}\int_{t_{0}}^{t_{\rm f}}d\tau\int_{t_{0}}^{t_{\rm f}}d\tau'~4\pi\eta_{\mathbf{x}}~g(\mathbf{x})~\Delta\dot{\phi}(\mathbf{x},\tau)\left[-2~\mathcal{D}_{\mathbf{x}}(\tau-\tau')~\Sigma\dot{\phi}(\mathbf{x},\tau')+\frac{i}{2}~\mathcal{N}_{\mathbf{x}}(\tau,\tau')~\Delta\dot{\phi}(\mathbf{x},\tau')\right]\nonumber\\
&=&\int d^{4}x\int d^{4}x'~\Delta\dot{\phi}(x)\left[-2~\mathcal{D}(x,x')~\Sigma\dot{\phi}(x')+\frac{i}{2}~\mathcal{N}(x,x')~\Delta\dot{\phi}(x')\right].
\label{FieldSIFCurrent}
\end{eqnarray}

Now, to continue the calculation as in the last Section, and to identify the noise and dissipation kernels of the present model, we integrate by parts in both time variables to obtain
an influence action depending on the sum and difference of the fields, instead of their time derivatives. Therefore, as in Ref. \cite{ArteagaBarrielTesis}, we obtain

\begin{eqnarray}
\widetilde{S}_{\rm IF}[\phi,\phi']&=&\int d\mathbf{x}\int_{t_{0}}^{t_{\rm f}}d\tau\int_{t_{0}}^{t_{\rm f}}d\tau'4\pi\eta_{\mathbf{x}} g(\mathbf{x})~\Delta\phi(\mathbf{x},\tau)\left[-2\partial_{\tau\tau'}^{2}\mathcal{D}_{\mathbf{x}}(\tau-\tau')\Sigma\phi(\mathbf{x},\tau')+\frac{i}{2}\partial_{\tau\tau'}^{2}\mathcal{N}_{\mathbf{x}}(\tau,\tau') \Delta\phi(\mathbf{x},\tau')\right].
\label{FieldSIFCurrentIntParts}
\end{eqnarray}

Since the dissipation kernel $\mathcal{D}$ involves the product of two distributions (because $\mathcal{D}(\tau-\tau')$ contains $\Theta(\tau-\tau')$ times an accompanying function of the times difference), the kernel is not well-defined  \cite{ArteagaBarrielTesis}. Differentiating twice the kernel, firstly with respect $\tau'$ and secondly respect to $\tau$, causes that

\begin{equation}
\partial_{\tau\tau'}^{2}\mathcal{D}_{\mathbf{x}}(\tau-\tau')=-\delta(\tau-\tau')~\dot{\mathcal{D}}_{\mathbf{x}}(\tau-\tau')-\ddot{\mathcal{D}}_{\mathbf{x}}(\tau-\tau'),
\label{DobleDerD}
\end{equation}
 where dots over the kernels represent time derivatives involving differentiation over the accompanying function of the times difference, avoiding the differentiation of the Heavyside function contained in the kernel. Without confusion, it is remarkable that in the first term the Dirac delta function comes from the differentiation of the Heavyside function, but the actual notation makes that the Heavyside function contained in $\dot{\mathcal{D}}_{\mathbf{x}}$ is superfluous and meaningless. On the other hand, we shall also note that we have exploited the fact that the dissipation kernel $\mathcal{D}$ depends on the times difference $\tau-\tau'$, which gives $\partial_{\tau'}\mathcal{D}=-\partial_{\tau}\mathcal{D}=-\dot{\mathcal{D}}$. On the other hand, this is unnecessary for the kernel $\mathcal{N}_{\mathbf{x}}$.

Inserting Eq.(\ref{DobleDerD}) into the influence action Eq.(\ref{FieldSIFCurrentIntParts}), by considering that from its definition $\dot{\mathcal{D}}_{\mathbf{x}}(0^{+})=\lambda_{0,\mathbf{x}}^{2}/2$ for the first term of Eq.(\ref{DobleDerD}), we clearly obtain

\begin{eqnarray}
\widetilde{S}_{\rm IF}[\phi,\phi']&=&\int d\mathbf{x}\int_{t_{0}}^{t_{\rm f}}d\tau~4\pi\eta_{\mathbf{x}}\lambda_{0,\mathbf{x}}^{2}~g(\mathbf{x})~\Delta\phi(\mathbf{x},\tau)~\Sigma\phi(\mathbf{x},\tau)\nonumber\\
&&+\int d\mathbf{x}\int_{t_{0}}^{t_{\rm f}}d\tau\int_{t_{0}}^{t_{\rm f}}d\tau'~4\pi\eta_{\mathbf{x}}~g(\mathbf{x})~\Delta\phi(\mathbf{x},\tau)\left[2~\ddot{\mathcal{D}}_{\mathbf{x}}(\tau-\tau')~\Sigma\phi(\mathbf{x},\tau')+\frac{i}{2}~\partial_{\tau\tau'}^{2}\mathcal{N}_{\mathbf{x}}(\tau,\tau')~\Delta\phi(\mathbf{x},\tau')\right]
\label{FieldSIFCurrentIntPartsFINAL}
\end{eqnarray}
where the first term is a finite renormalization position-dependent mass term for the scalar field which will be meaningless in the determination of the Green function as will see in next Sections. These renormalization mass terms also appears in the QBM theory, but in general they are divergent due to the fact that the bath is a set of infinite harmonic oscillators, each one contributing to the mass renormalization. In our case, the field is coupled in each space point $\mathbf{x}$ to an unique harmonic oscillator represented by the polarization degree of freedom located at $\mathbf{x}$, so the renormalization term is only one, and then, it is finite.

It is worth noting that from the second term, we shall call current-dissipation kernel and current-noise kernel, to the derivatives of the dissipation and noise kernels of the bilinear model, i. e., the current-dissipation kernel is $-\ddot{\mathcal{D}}_{\mathbf{x}}$ while the current-noise kernel is $\partial_{\tau\tau'}^{2}\mathcal{N}_{\mathbf{x}}$. From this Section and so on, and to avoid confusion, we shall use the prefixed 'current' for the kernels referring to the current-type model, keeping the terms dissipation and noise kernels for $\mathcal{D}_{\mathbf{x}}$ and $\mathcal{N}_{\mathbf{x}}$ respectively.

All in all, and having written the influence action of Eq.(\ref{FieldSIFCurrentIntPartsFINAL}) formally as a renormalization mass term plus a non-local term (identical to Eq.(\ref{FieldSIF}) but with different kernels), we can go back and resume the procedure done for the bilinear coupling in the last Section.

Despite the renormalization mass term, the CTP functional integral over the field variables can be done as in the last Section. Therefore, the generating functional results formally identical to Eq.(\ref{GeneratingFunctionalHighTFINAL}). However, in the present case, the current-noise and current-dissipation kernels are different, so the first one will define the contribution due to the matter fluctuations, while the second one will contribute to the definition of the retarded Green function by appearing in the field's semiclassical equation obtained from the CTP effective action for the current-type model. This equation can be easily derived as Eq.(\ref{EqMotionPhi}), obtaining

\begin{equation}
\partial_{\mu}\partial^{\mu}\phi+4\pi\eta_{\mathbf{x}}\lambda_{0,\mathbf{x}}^{2}~g(\mathbf{x})~\phi(\mathbf{x},t)+8\pi\eta_{\mathbf{x}}~g(\mathbf{x})\int_{t_{0}}^{t} d\tau~\ddot{\mathcal{D}}_{\mathbf{x}}(t-\tau)~\phi(\mathbf{x},\tau)=0,
\label{EqMotionPhiCurrent}
\end{equation}
where the scalar field has a well-defined (positive) position-dependent mass $2\sqrt{\pi\eta_{\mathbf{x}}}~|\lambda_{0,\mathbf{x}}|$ in every point $\mathbf{x}$ where there is material (so $g(\mathbf{x})=1$), while it is massless in the free regions. This last equation is in agreement with the one obtained from a canonical quantization scheme (see Ref. \cite{LombiMazziRL} for example) and it is in fact its generalization.

\section{Energy-Momentum Tensor and Field Correlation}\label{EMTFC}

At this point, we have obtained the field CTP generating functional for both coupling models after tracing out all the material degrees of freedom (polarization plus thermal baths). Then, we are interested in evaluating the expectation value of the symmetric energy-momentum tensor operator $\langle\widehat{T}_{\mu\nu}\rangle$, which gives the energy density and radiation pressure associated to the field, it is defined by \cite{GreiRein, Ramond}:

\begin{equation}
\widehat{T}_{\mu\nu}(x_{1})\equiv-\eta_{\mu\nu}~\frac{1}{2}~\partial_{\gamma}\widehat{\phi}(x_{1})~\partial^{\gamma}\widehat{\phi}(x_{1})+\partial_{\mu}\widehat{\phi}(x_{1})~\partial_{\nu}\widehat{\phi}(x_{1}),
\end{equation}
where $\eta_{\mu\nu}$ is the Minkowski metric ($\eta_{00}=-\eta_{ii}=1$ for the non-vanishing elements).

We can proceed through the point splitting technique, employing the field correlation function as

\begin{equation}
\Big\langle\widehat{T}_{\mu\nu}(x_{1})\Big\rangle=\lim_{x_{2}\rightarrow x_{1}}\left(-\eta_{\mu\nu}~\frac{1}{2}~\partial_{\gamma_{1}}\partial^{\gamma_{2}}+\partial_{\mu_{1}}\partial_{\nu_{2}}\right)\Big\langle\widehat{\phi}(x_{1})\widehat{\phi}(x_{2})\Big\rangle,
\end{equation}
where the notation implies $\partial_{\gamma_{1}}\partial^{\gamma_{2}}\equiv\partial_{t_{1}}\partial_{t_{2}}-\nabla_{1}\cdot\nabla_{2}$ and so on for $\partial_{\mu_{1}}\partial_{\nu_{2}}$.

Therefore, we need the field correlation function to know the expectation value of every energy-momentum tensor component. In fact, we need the correlation to be finite, so we have to insert a regularized expression of the correlation function. From the generating functional in Eq.(\ref{GeneratingFunctionalHighTFINAL}), this is straightforward \cite{CalHu}. We will evaluate the field correlation in two different points $x_{1}$ and $x_{2}$ (having no specific relation between the points because they are in different branches of the CTP). Then,
we have four alternatives depending on the relation between $x_{1}$ and $x_{2}$, however in the coincidence limit this is not relevant,

\begin{equation}
\Big\langle\widehat{\phi}(x_{1})\widehat{\phi}(x_{2})\Big\rangle=\frac{\delta^{2}Z}{\delta J'(x_{1})\delta J(x_{2})}\Big|_{J=J'=0}.
\end{equation}

Because the generating functional has a simple form in Eq.(\ref{GeneratingFunctionalHighTFINAL}), we can easily compute its functional derivatives, taking advantage of the symmetry kernel's properties, to obtain:

\begin{equation}
\Big\langle\widehat{\phi}(x_{1})\widehat{\phi}(x_{2})\Big\rangle=\mathcal{A}(x_{1},x_{2})+\mathcal{B}(x_{1},x_{2})+\int d^{4}x\int d^{4}x'~\mathcal{G}_{\rm Ret}(x_{1},x)~\mathcal{N}(x,x')~\mathcal{G}_{\rm Ret}(x_{2},x')+\frac{1}{2}~\mathcal{G}_{\rm Jordan}(x_{1},x_{2}),
\label{FieldCorrelationHighT}
\end{equation}
where $\mathcal{G}_{\rm Jordan}(x_{1},x_{2})\equiv i\left(\mathcal{G}_{\rm Ret}(x_{2},x_{1})-\mathcal{G}_{\rm Ret}(x_{1},x_{2})\right)$ is the Jordan propagator \cite{CalHu}. Then, the kernels are the ones in Eqs.(\ref{FieldSIF}), (\ref{KernelAHighT}) and (\ref{KernelBHighT}) for the case of the bilinear model, being the retarded Green function defined from semiclassical equation of motion on Eq.(\ref{EqMotionPhi}). On the other hand, to obtain the result for the current-type model we have to take into account that the retarded Green function is defined from the corresponding semiclassical equation of motion for the field in this model, given by Eq.(\ref{EqMotionPhiCurrent}), but the formal expressions for the kernels $\mathcal{A}$ and $\mathcal{B}$ are unchanged. To finish, we have to replace the noise kernel $\mathcal{N}$ on Eq.(\ref{FieldCorrelationHighT}) by the current-noise kernel $\partial_{\tau\tau'}^{2}\mathcal{N}$ associated to the field's influence action for this model of Eq.(\ref{FieldSIFCurrentIntPartsFINAL}). All in all, a smart and compact notation can be achieved by including a parameter $\alpha$ encompassing both models, therefore, we can write the generalized noise kernel as $\partial_{\tau\tau'}^{2\alpha}\mathcal{N}$, with $\alpha=0,1$ for the bilinear and current-type model respectively.

It is worth noting that the correlation function in Eq.(\ref{FieldCorrelationHighT}) corresponds to the Whightman function for the field in this open system. In fact, considering that $\mathcal{G}_{\rm Ret}$ is real, as it is written is clear that the correlation is a complex quantity, and its imaginary part is given by $\mathcal{G}_{\rm Jordan}$ whereas that the real part is formed by the others three terms. If we want to match the Whightman propagator with the typical relations for propagators, the Hadamard propagator is given by

\begin{equation}
\mathcal{G}_{\rm H}(x_{1},x_{2})\equiv 2\left[\mathcal{A}(x_{1},x_{2})+\mathcal{B}(x_{1},x_{2})+\int d^{4}x\int d^{4}x'~\mathcal{G}_{\rm Ret}(x_{1},x)~\partial_{\tau\tau'}^{2\alpha}\left[\mathcal{N}(x,x')\right]~\mathcal{G}_{\rm Ret}(x_{2},x')\right].
\label{HadamardPropagator}
\end{equation}

On the other hand, we want to calculate energy-momentum tensor expectation values, so the result must be real. This apparently seems not to be the case because the correlation function is complex and its imaginary part is given by $\mathcal{G}_{\rm Jordan}$. However, to compute the expectation values, we will have to derive in a symmetric way in both coordinates $x_{1,2}$ and then we will have to calculate the coincidence limit when $x_{2}\rightarrow x_{1}$. Due to the definition of the Jordan propagator, this operation (symmetric derivation plus the coincidence limit) makes that the contribution vanishes. Therefore, the expectation values turns out to be, effectively, real numbers, as is expected.

Finally, the expectation value of the energy-momentum tensor can be written as

\begin{equation}
\Big\langle\widehat{T}_{\mu\nu}(x_{1})\Big\rangle=\frac{1}{2}\lim_{x_{2}\rightarrow x_{1}}\left(-\eta_{\mu\nu}~\frac{1}{2}~\partial_{\gamma_{1}}\partial^{\gamma_{2}}+\partial_{\mu_{1}}\partial_{\nu_{2}}\right)\mathcal{G}_{\rm H}(x_{1},x_{2}),
\label{TmunuExpValue}
\end{equation}
where the Hadamard propagator must be a well-defined (non-divergent) propagator.

It is important to note that all the full non-equilibrium dynamics, both time evolution and thermodynamical non-equilibrium, is contained in this result.

Due to the structure of the noise kernel $\partial_{\tau\tau'}^{2\alpha}\mathcal{N}$ for each model, this term accounts for the influence generated by the material (polarization degrees of freedom plus thermal baths), from the initial time when the interaction with the field is turn on, describing the relaxation process of the material forming the contours. Both parts, polarization degrees of freedom and thermal baths, can have different initial temperature, having thermal non-equilibrium. In fact, each volume element on the material can have its own properties.

On the other hand, there are two terms proportionals to the field initial temperature, which are the kernels $\mathcal{A}$ and $\mathcal{B}$ (and their derivatives in the contribution of the expectation values). Those terms account for the dynamical evolution and change of the field in the presence of the material contours, when the interaction is turned on. Therefore, those terms must be entirely related to the modified normal modes that appear in a (steady situation) canonical quantization scheme as a vacuum contribution \cite{LombiMazziRL,Dorota1992}.

\section{Non-equilibrium behaviour for different configurations of the composite system}\label{NEBFDCOTCS}

\subsection{$0+1$ Field}\label{0+1F}

As a first example, we consider the case of a scalar field in $0+1$ dimensions, i. e., we take the field $\phi$ as a quantum harmonic oscillator degree of freedom of unit mass. Therefore, to adapt our results to this situation, a few changes are needed. In this case the spatial notion is erased, and the volume element concept is meaningless, so the composite system is an harmonic oscillator ($0+1$ field) coupled to another one (polarization degree of freedom) which is also coupled to a set of harmonic oscillators (thermal bath).

The spatial label $\mathbf{x}$ will be unnecessary and the quantum degree of freedom will be characterized by a frequency $\Omega$ (which plays the role that the spatial derivative has on the field in $n+1$ dimensions, with $n>0$). The initial action for the field (Eq.(\ref{FreeFieldAction})) must be replaced by the straightforward expression

\begin{equation}
S_{0}[\phi]=\int d\mathbf{x}\int_{t_{0}}^{t_{\rm f}}d\tau~\frac{1}{2}~\partial_{\mu}\phi~\partial^{\mu}\phi\longrightarrow\int_{t_{0}}^{t_{\rm f}}
d\tau~\frac{1}{2}~\left[\left(\frac{d\phi}{d\tau}\right)^{2}-\Omega^{2}~\phi(\tau)\right].
\end{equation}

Eqs.(\ref{FreePolAction}) - (\ref{IntPolBathHOAction}) can also be simply adapted by discarding the spatial integrals, labels, the density $\eta$ (together with the factor $4\pi$) and the distribution function $g$ in all the actions. All the integrations and traces can be performed in the same way without further modifications until the functional integration over the field. The influence action in Eq.(\ref{FieldSIF}) still remains valid. In this way, the generating functional from Eq.(\ref{GeneratingFunctional}) to Eq.(\ref{GeneratingFunctionalNOFinal}) is formally the same. However, in this case, the calculation of the first factor, that involves the initial state of the $0+1$ field $\phi(t)$, implies a Wigner function and not a functional, so the factor results from the same formal calculation done for the polarization degree of freedom $r$ in Eq.(\ref{FirstFactorPol}), but the kernel obtained is clearly different. Therefore, we have for the both models ($\alpha=0,1$):

\begin{eqnarray}
Z[J,J']&=&e^{-\frac{1}{2}\int_{t_{0}}^{t_{\rm f}}d\tau\int_{t_{0}}^{t_{\rm f}}ds~J_{\Delta}(\tau)\left[\mathcal{A}(\tau,s)+\mathcal{B}(\tau,s)\right]J_{\Delta}(s)}~e^{-\frac{1}{2}\int_{t_{0}}^{t_{\rm f}}d\tau\int_{t_{0}}^{t_{\rm f}}d\tau'\int_{t_{0}}^{t_{\rm f}}ds'\int_{t_{0}}^{t_{\rm f}}ds~J_{\Delta}(\tau)~\mathcal{G}_{\rm Ret}^{\Omega}(\tau,\tau')~\partial_{\tau's'}^{2\alpha}\left[\mathcal{N}(\tau',s')\right]~\mathcal{G}_{\rm Ret}^{\Omega}(s,s')~J_{\Delta}(s)}\nonumber\\
&\times &~e^{-i\int_{t_{0}}^{t_{\rm f}}d\tau\int_{t_{0}}^{t_{\rm f}}ds~J_{\Delta}(\tau)~\mathcal{G}_{\rm Ret}^{\Omega}(\tau,s)~J_{\Sigma}(s)},
\label{GeneratingFunctional0+1}
\end{eqnarray}
where the sum of the kernels $\mathcal{A}$ and $\mathcal{B}$ results from the ordinary integration over the initial values of the field $\phi_{0}\equiv\phi(t_{0})$ and $\Pi_{0}\equiv\Pi(t_{0})$ and it is of the form of Eq.(\ref{PhiNoiseKernelR}) (indeed, the high temperature limit of this expression has exactly the form of Eq.(\ref{PhiNoiseKernelR}) by discarding it spatial features),

\begin{equation}
\mathcal{A}(\tau,s)+\mathcal{B}(\tau,s)\equiv\frac{1}{2\Omega}~\coth\left(\frac{\beta_{\phi}\Omega}{2}\right)\left[\Omega^{2}~\mathcal{G}_{\rm Ret}^{\Omega}(\tau-t_{0})~\mathcal{G}_{\rm Ret}^{\Omega}(s-t_{0})+\dot{\mathcal{G}}_{\rm Ret}^{\Omega}(\tau-t_{0})~\dot{\mathcal{G}}_{\rm Ret}^{\Omega}(s-t_{0})\right],
\end{equation}
where $\mathcal{G}_{\rm Ret}^{\Omega}$ is the retarded Green function (which is a function of the time difference, i. e., $\mathcal{G}_{\rm Ret}^{\Omega}(t,s)\equiv\mathcal{G}_{\rm Ret}^{\Omega}(t-s)$, as we can infer from its equation of motion), associated to Eqs. (\ref{EqMotionPhi}) and (\ref{EqMotionPhiCurrent}) for each model respectively in the $0+1$ case, which can be written together as

\begin{eqnarray}
\frac{d^{2}\phi}{dt^{2}}+(\Omega^{2}+\alpha~\lambda_{0}^{2})~\phi(t)-(-1)^{\alpha}~2\int_{t_{0}}^{t}d\tau~\partial_{tt}^{2\alpha}\left[\mathcal{D}(t-\tau)\right]~\phi(\tau)=0,
\label{EqMotionPhi0+1}
\end{eqnarray}
where in this case, the (finite) mass term presents as a frequency renormalization term, and $\partial_{tt}^{2\alpha}\left[\mathcal{D}(t-\tau)\right]$ is the generalized dissipation kernel.

Therefore, the $0+1$ counterpart of the Hadamard propagator of Eq.(\ref{HadamardPropagator}) is

\begin{eqnarray}
\mathcal{G}_{\rm H}^{\Omega}(t_{1},t_{2})&\equiv& \frac{1}{\Omega}~\coth\left(\frac{\beta_{\phi}\Omega}{2}\right)\left[\Omega^{2}~\mathcal{G}_{\rm Ret}^{\Omega}(t_{1}-t_{0})~\mathcal{G}_{\rm Ret}^{\Omega}(t_{2}-t_{0})+\dot{\mathcal{G}}_{\rm Ret}^{\Omega}(t_{1}-t_{0})~\dot{\mathcal{G}}_{\rm Ret}^{\Omega}(t_{2}-t_{0})\right]\nonumber\\
&&+~2\int_{t_{0}}^{t_{\rm f}}d\tau\int_{t_{0}}^{t_{\rm f}}d\tau'~\mathcal{G}_{\rm Ret}^{\Omega}(t_{1}-\tau)~\partial_{\tau\tau'}^{2\alpha}\left[\mathcal{N}(\tau,\tau')\right]~\mathcal{G}_{\rm Ret}^{\Omega}(t_{2}-\tau'),
\label{HadamardPropagator0+1}
\end{eqnarray}
with the noise kernel in two contributions $\mathcal{N}(\tau,\tau')=\mathcal{N}_{B}(\tau,\tau')+\mathcal{N}_{r}(\tau,\tau')$, each one characterized by its own temperature $\beta_{r,B}$, given in Eqs.(\ref{PhiNoiseKernelB}) and (\ref{PhiNoiseKernelR}) respectively. In fact, we can correspond the temperature value associated to the term to the contribution of that part of the total system, i. e., the terms carrying the field's temperature $\beta_{\phi}$ are associated to the proper (influenced) system contribution, while each part of the noise kernel $\mathcal{N}$ has one term associated to the polarization degree of freedom (denoted by containing the temperature $\beta_{r}$) and another one associated to the bath (denoted by containing the temperature $\beta_{B}$).

It is clear now, that the energy-momentum tensor is simply the energy of the $0+1$ field, where the evolution of the expectation value can be easily written in terms of the Hadamard propagator, as it happens in Eq.(\ref{TmunuExpValue}):

\begin{equation}
\Big\langle E(t_{1})\Big\rangle\equiv\frac{1}{2}\lim_{t_{2}\rightarrow t_{1}}\left(\frac{\partial}{\partial t_{1}}\frac{\partial}{\partial t_{2}}+\Omega^{2}\right)\mathcal{G}_{\rm H}^{\Omega}(t_{1},t_{2}).
\label{EnergyExpValue}
\end{equation}

Finally, we have written the mean value of the energy as a function of time, from the initial conditions for the composite system. It is clear that the dynamic depends on the retarded Green functions $\mathcal{G}_{\rm Ret}^{\Omega}, G_{\rm Ret}$ (where $G_{\rm Ret}$ is contained in the field's noise kernels of each model through Eqs. (\ref{PhiNoiseKernelB}) and (\ref{PhiNoiseKernelR})), from each part of the system and the QBM noise kernel $N_{\rm QBM}$ (which depends on the type of bath we are considering) after tracing out the degrees of freedom that influences its dynamics. Since we are interested in the field dynamics, the traces are performed taking the field $\phi$ as the system and doing them in sequential steps: the partial traces over each part of the complex environment formed by the polarization degree of freedom $r$ and the bath $\{q_{n}\}$ \cite{FeynHibbs}.

Therefore, the transient time behavior of the energy expectation value and its relaxation to a steady state, will depend on the fluctuations of each part of the environment, through the noise kernels, and how the system evolves to the steady situation, depends on its own Green function $\mathcal{G}_{\rm Ret}^{\Omega}$, as it is clear from Eq.(\ref{HadamardPropagator0+1}).

Then, for the long-time limit ($t_{0}\rightarrow-\infty$), we need to know how is the long-time behavior of each retarded Green function $\mathcal{G}_{\rm Ret}^{\Omega}, G_{\rm Ret}$. Thus, we must focus on the specific Green functions that we have in our system, which are determined by each equation of motion we have obtained at each stage of the tracing.

The retarded Green function for the polarization degree of freedom $r$ is determined by the equation of motion of the polarization degree of freedom, Eq.(\ref{EqMotionR}). The associated equation for the Green function $G_{\rm Ret}$ can be solved by Laplace transforming the equation subjected to the initial conditions $G_{\rm Ret}(0)=0,\dot{G}_{\rm Ret}(0)=1$ (see Ref. \cite{BreuerPett}). It is straightforward to prove that, for every type of bath, the Laplace transform of the retarded Green function is given by

\begin{equation}
\widetilde{G}_{\rm Ret}(z)=\frac{1}{\left(z^{2}+\omega^{2}-2~\widetilde{D}_{\rm QBM}(z)\right)},
\label{RetGreenFunctionQBM}
\end{equation}
where $\widetilde{D}_{\rm QBM}$ is the Laplace transform of the QBM's dissipation kernel contained in $S_{QBM}$ \cite{CaldeLegg, HuPazZhang}.

The analyticity properties of the the Laplace transform $\widetilde{G}_{\rm Ret}$ and the location of its poles define the time evolution and the asymptotic behavior of the Green function $G_{\rm Ret}$. In this direction, causality implies, by Cauchy's theorem, that the poles of $\widetilde{G}_{\rm Ret}$ should be located in the left-half of the complex $z$-plane, i. e., the poles' real parts must be negative or zero. Assuming that $\omega\neq 0$ and that the bath modeled includes a cutoff function in frequencies (see \cite{BreuerPett}), considering the discussion given in Ref. \cite{LombiMazziRL}, which results that all the poles are simple and have negative real parts. Through the Mellin's formula and the Residue theorem to retransform to the time dependent function \cite{Schiff}, we easily obtain that, formally, the Green function reads

\begin{equation}
G_{\rm Ret}(t)=\Theta(t)\sum_{j}Res\left[\widetilde{G}_{\rm Ret}(z),z_{j}\right]~e^{z_{j}t}.
\end{equation}

Since $Re[z_{j}]<0$, it is clear that in the long-time limit, when $t_{0}\rightarrow-\infty$, we have $G_{\rm Ret}(t-t_{0})\rightarrow 0$ and also the same for its time derivatives.

Indeed, this asymptotic behavior defines the long-time contribution of the polarization degree of freedom to the field's energy density at the steady situation. Since the Green function goes to zero, we have also that the part of the field's noise kernels, directly associated to the polarization degree of freedom $\mathcal{N}_{r}$, goes to zero. This means that the polarization degree of freedom do not contribute through its thermal state to the energy at the steady situation in none of the two coupling models.  Although the dependence on the temperature $\beta_{r}$ is erased in the long-time regime (due to the asymptotic decay of the retarded Green function $G_{\rm Ret}(t-t_{0})$), this function also appears in the bath's contribution $\mathcal{N}_{B}$. That term is characterized, of course, by the bath's temperature $\beta_{B}$.

All in all, in the long-time limit ($t_{0}\rightarrow-\infty$) we have that the (generalized) noise kernel contribution (polarization degree of freedom plus bath) in Eq.(\ref{HadamardPropagator0+1}) results

\begin{equation}
\int_{t_{0}}^{t_{\rm f}}d\tau\int_{t_{0}}^{t_{\rm f}}d\tau'~\mathcal{G}_{\rm Ret}^{\Omega}(t_{1}-\tau)~\partial_{\tau\tau'}^{2\alpha}\left[\mathcal{N}(\tau,\tau')\right]~\mathcal{G}_{\rm Ret}^{\Omega}(t_{2}-\tau')\longrightarrow\int_{-\infty}^{t_{\rm f}}d\tau\int_{-\infty}^{t_{\rm f}}d\tau'~\mathcal{G}_{\rm Ret}^{\Omega}(t_{1}-\tau)~\partial_{\tau\tau'}^{2\alpha}\left[\mathcal{N}_{B}(\tau,\tau')\right]~\mathcal{G}_{\rm Ret}^{\Omega}(t_{2}-\tau')
\end{equation}
where the QBM noise kernel does not depends on $t_{0}$ so it makes that the bath contribution do not vanish in the steady situation.

Finally, we have to analyze the behavior of the contribution associated to the proper field-system. Thus we have to study the retarded Green function $\mathcal{G}_{\rm Ret}^{\Omega}$. We then proceed as in the polarization degree of freedom case, for studying $G_{\rm Ret}$ by considering the same initial conditions ($\mathcal{G}_{\rm Ret}^{\Omega}(0)=0,\dot{\mathcal{G}}_{\rm Ret}^{\Omega}(0)=-1$). From the equation of motion for the Green function $\mathcal{G}_{\rm Ret}^{\Omega}$, associated to the field in both models, Eq.(\ref{EqMotionPhi0+1}), we can easily obtain an analogous expression as in the first case for the Laplace transform

\begin{eqnarray}
\widetilde{\mathcal{G}}_{\rm Ret}^{\Omega}(z)&=&\frac{-1}{\left(z^{2}+\Omega^{2}-\lambda_{0}^{2}~(-z^{2})^{\alpha}~\widetilde{G}_{\rm Ret}(z)\right)},
\end{eqnarray}
where it is worth noting that this compact expression is due to the fact that the renormalization (mass) frequency term cancels out with a term coming from the derivative of the dissipation kernel $\mathcal{D}$ at the initial time.

Analyticity properties of this Laplace transform define the asymptotic behavior of the proper contribution of the field. For $\lambda_{0},\Omega,\omega\neq 0$ and an Ohmic bath, it is easy to show that the Laplace transform for both models has four simple poles with negative real parts, verifying the causality requirement. We assume that the general case gives the same features and the poles are simple and have negative real parts. From this, in the time domain, it follows that

\begin{equation}
\mathcal{G}_{\rm Ret}^{\Omega}(t)=\Theta(t)\sum_{l}Res\left[\widetilde{\mathcal{G}}_{\rm Ret}^{\Omega}(z),z_{l}\right]~e^{z_{l}t}.
\end{equation}

Therefore, since $Re[z_{l}]<0$, we clearly have in the long-time limit ($t_{0}\rightarrow-\infty$) that $\mathcal{G}_{\rm Ret}^{\Omega}(t-t_{0})\rightarrow 0$ and also the same for its time derivatives.

The long-time limit of the Hadamard propagator $\mathcal{G}_{\rm H}^{\Omega}$ is given only by the bath's long-time contribution:

\begin{eqnarray}
\mathcal{G}_{\rm H}^{\Omega}(t_{1},t_{2})\rightarrow 2\int_{-\infty}^{t_{\rm f}}d\tau\int_{-\infty}^{t_{\rm f}}d\tau'~\mathcal{G}_{\rm Ret}^{\Omega}(t_{1}-\tau)~\mathcal{N}_{B}(\tau,\tau')~\mathcal{G}_{\rm Ret}^{\Omega}(t_{2}-\tau'),
\label{HadamardPropagator0+1}
\end{eqnarray}
corresponding to the steady situation with the bath's fluctuation at temperature $\beta_{B}$, as the fluctuation-dissipation theorem asserts.

Finally, summarizing thus section, for a $0+1$ field in both types of coupling models, the energy density at the steady situation, have only contributions from the bath, while the polarization degree of freedom and the proper field contributions go to zero through the time evolution.

Now, let see how these calculations apply for the case of a field in $n+1$ dimensions with an homogeneous material all over the space.

\subsection{Field In Infinite Material}\label{FIIM}

Let us now consider a scalar field in $n+1$ dimensions (with $n\neq 0$) with no boundaries, i. e., this is the case of an homogeneous material that appears all over the space at the initial time $t_{0}$. In this situation, $g(\mathbf{x})\equiv 1$ for every $\mathbf{x}$ and we have to eliminate the spatial label due to the homogeneity of the problem. Then, Eqs.(\ref{EqMotionPhi}) and (\ref{EqMotionPhiCurrent}) can be written together through its generalized form as

\begin{eqnarray}
\partial_{\mu}\partial^{\mu}\phi+4\pi\eta\lambda_{0}^{2}~\alpha~\phi-(-1)^{\alpha}8\pi\eta\int_{t_{0}}^{t}d\tau~\partial_{tt}^{2\alpha}\left[\mathcal{D}(t-\tau)\right]~\phi(\mathbf{x},\tau)&=&0,
\label{EqMotionPhiHomogeneousAllSpace}
\end{eqnarray}
which is basically a wave-type equation for the field in a dissipative media.

Therefore, the associated equation for the retarded Green function $\mathcal{G}_{\rm Ret}$ is straightforward and it is subjected to the typical wave equation initial conditions

\begin{equation}
\mathcal{G}_{\rm Ret}(\mathbf{x},\mathbf{x}',0)=0~~~~~,~~~~~\dot{\mathcal{G}}_{\rm Ret}(\mathbf{x},\mathbf{x}',0)=-\delta(\mathbf{x}-\mathbf{x}').
\label{InitialConditionsHomogeneous}
\end{equation}

Due to the translational symmetry of the problem, $\mathcal{G}_{\rm Ret}(\mathbf{x},\mathbf{x}',t)=\mathcal{G}_{\rm Ret}(\mathbf{x}-\mathbf{x}',t)$, the Fourier transform satisfies

\begin{eqnarray}
\partial_{tt}^{2}\overline{\mathcal{G}}_{\rm Ret}(\mathbf{k},t)+(k^{2}+4\pi\eta\lambda_{0}^{2}~\alpha)~\overline{\mathcal{G}}_{\rm Ret}(\mathbf{k},t)-(-1)^{\alpha}8\pi\eta\int_{0}^{t}d\tau~\partial_{tt}^{2\alpha}\left[\mathcal{D}(t-\tau)\right]~\overline{\mathcal{G}}_{\rm Ret}(\mathbf{k},\tau)=0,
\label{EqMotionPhiHomogeneousAllSpaceFourier}
\end{eqnarray}
 where, as in the last section, $\mathcal{D}(t-\tau)=\lambda_{0}^{2}~G_{\rm Ret}(t-\tau)$, $k\equiv|\mathbf{k}|$,  and the initial conditions are

\begin{equation}
\overline{\mathcal{G}}_{\rm Ret}(\mathbf{k},0)=0~~~~~,~~~~~\dot{\overline{\mathcal{G}}}_{\rm Ret}(\mathbf{k},0)=-1.
\label{InitialConditionsHomogeneousFourier}
\end{equation}

Eqs.(\ref{EqMotionPhiHomogeneousAllSpaceFourier}) and (\ref{InitialConditionsHomogeneousFourier}) are equivalent to the field equation and initial conditions for the retarded Green function for the $0+1$ field, i.e., each field mode behaves as a $0+1$ field of natural frequency $k$ and the dynamics are equivalent. Then, the Fourier transform of the retarded Green function is closely related to the retarded Green function in the last Section, in fact, we have,

\begin{equation}
\overline{\mathcal{G}}_{\rm Ret}(\mathbf{k},t)\equiv\overline{\mathcal{G}}_{\rm Ret}^{k}(t),
\end{equation}
where $\mathcal{G}_{\rm Ret}^{k}$ is the retarded function of a $0+1$ field of frequency $k$.

We can write

\begin{equation}
\mathcal{G}_{\rm Ret}(\mathbf{x}-\mathbf{x}',t)=\int\frac{d\mathbf{k}}{(2\pi)^{3}}~e^{-i\mathbf{k}\cdot(\mathbf{x}-\mathbf{x}')}~\mathcal{G}_{\rm Ret}^{k}(t).
\end{equation}

In order to study the behavior of the contributions to the expectation value of the energy-momentum tensor  $\langle\widehat{T}_{\mu\nu}\rangle$ in Eq.(\ref{TmunuExpValue}),  let us firstly consider the contributions of the polarization degrees of freedom and the thermal baths in each point $\mathbf{x}$ in the last term of Eq.(\ref{HadamardPropagator}).  Since we are considering an homogeneous material with all the polarization degrees of freedom having the same temperature $\beta_{r}$ and the same for the baths in each point with $\beta_{B}$ (note that this does not means thermal equilibrium because each part of the material can have different temperatures, i. e., we can still have the situation in which $\beta_{r}\neq\beta_{B}$). In the present case $\mathcal{N}(x,x')=4\pi\eta~\delta(\mathbf{x}-\mathbf{x}')~\mathcal{N}(\tau,\tau')$.

If we use the Fourier representation of $\mathcal{G}_{\rm Ret}$ to write the last term of Eq.(\ref{HadamardPropagator}), it is straightforward that

\begin{eqnarray}
\int d^{4}x\int d^{4}x'&\mathcal{G}_{\rm Ret}(x_{1},x)&~\partial_{\tau\tau'}^{2\alpha}\left[\mathcal{N}(x,x')\right]\mathcal{G}_{\rm Ret}(x_{2},x')\nonumber\\
&=4\pi\eta&\int\frac{d\mathbf{k}}{(2\pi)^{3}}e^{-i\mathbf{k}\cdot(\mathbf{x}_{1}-\mathbf{x}_{2})}\int_{t_{0}}^{t_{\rm f}}d\tau\int_{t_{0}}^{t_{\rm f}}d\tau'~\overline{\mathcal{G}}_{\rm Ret}^{k}(t_{1}-\tau)~\partial_{\tau\tau'}^{2\alpha}\left[\mathcal{N}(\tau,\tau')\right]\overline{\mathcal{G}}_{\rm Ret}^{k}(t_{2}-\tau'),
\end{eqnarray}
where it is remarkable that both integrals over $\tau$ and $\tau'$, and the integrand are exactly one half of the last term in Eq.(\ref{HadamardPropagator0+1}), the contribution of the polarization degree of freedom and the bath in the last Section, i. e., for the $0+1$ field of frequency $\Omega$. This is clear because, as we have inferred from the equation for the Fourier transformed Green function, each field $\mathbf{k}$-mode is matched to a $0+1$ field of frequency $k=|\mathbf{k}|$.

Then, we have for each field mode the same time evolution as for a $0+1$ field of natural frequency $k$ in any coupling model. Considering the analysis done in the last Section about the Green function $G_{\rm Ret}$, we can easily conclude that the long-time regime ($t_{0}\rightarrow-\infty$) of this contribution is given by

\begin{eqnarray}
\int d^{4}x\int d^{4}x'~&\mathcal{G}_{\rm Ret}(x_{1},x)&~\partial_{\tau\tau'}^{2\alpha}\left[\mathcal{N}(x,x')\right]~\mathcal{G}_{\rm Ret}(x_{2},x')\longrightarrow\nonumber\\
&\longrightarrow 4\pi\eta&\int\frac{d\mathbf{k}}{(2\pi)^{3}}e^{-i\mathbf{k}\cdot(\mathbf{x}_{1}-\mathbf{x}_{2})}\int_{-\infty}^{t_{\rm f}}d\tau\int_{-\infty}^{t_{\rm f}}d\tau'~\overline{\mathcal{G}}_{\rm Ret}^{k}(t_{1}-\tau)~\partial_{\tau\tau'}^{2\alpha}\left[\mathcal{N}_{B}(\tau,\tau')\right]~\overline{\mathcal{G}}_{\rm Ret}^{k}(t_{2}-\tau'),
\end{eqnarray}
where, as in the last Section, we have that the polarization degrees of freedom do not contribute to the steady situation of the $n+1$ field in an homogeneous material.

On the other hand, for the proper contribution of the field, contained in the kernels $\mathcal{A}$ and $\mathcal{B}$ of Eqs. (\ref{KernelAHighT}) and (\ref{KernelBHighT}), we can again exploit the Fourier representation

\begin{equation}
\mathcal{A}(x_{1},x_{2})+\mathcal{B}(x_{1},x_{2})=\int\frac{d\mathbf{k}}{(2\pi)^{3}}~e^{-i\mathbf{k}\cdot(\mathbf{x}_{1}-\mathbf{x}_{2})}\left[\frac{1}{\beta_{\phi}}~\overline{\mathcal{G}}_{\rm Ret}^{k}(t_{1}-t_{0})~\overline{\mathcal{G}}_{\rm Ret}^{k}(t_{2}-t_{0})+\overline{K}(k)~\dot{\overline{\mathcal{G}}}_{\rm Ret}^{k}(t_{1}-t_{0})~\dot{\overline{\mathcal{G}}}_{\rm Ret}^{k}(t_{2}-t_{0})\right].
\end{equation}

Therefore, considering the analysis done in the last Section for the retarded Green function $\mathcal{G}_{\rm Ret}^{\Omega}$, in the long-time limit ($t_{0}\rightarrow-\infty$) the Fourier transform of the retarded Green function vanishes, i. e., $\overline{\mathcal{G}}_{\rm Ret}^{k}(t-t_{0})\rightarrow 0$; and this makes that also the proper contribution vanishes at the steady situation.

All in all, as in the $0+1$ field, the long-time regime is defined by the bath contribution to the Hadamard propagator, and it is expected to satisfy the fluctuation-dissipation theorem in the steady situation by both coupling models

\begin{eqnarray}
\mathcal{G}_{\rm H}(x_{1},x_{2})\rightarrow 8\pi\eta\int\frac{d\mathbf{k}}{(2\pi)^{3}}~e^{-i\mathbf{k}\cdot(\mathbf{x}_{1}-\mathbf{x}_{2})}\int_{-\infty}^{t_{\rm f}}d\tau\int_{-\infty}^{t_{\rm f}}d\tau'~\overline{\mathcal{G}}_{\rm Ret}^{k}(t_{1}-\tau)~\partial_{\tau\tau'}^{2\alpha}\left[\mathcal{N}_{B}(\tau,\tau')\right]~\overline{\mathcal{G}}_{\rm Ret}^{k}(t_{2}-\tau').
\end{eqnarray}

Finally, the energy density at the steady situation will also depend only on the baths fluctuations in the long-time regime for anyone of the coupling models. This conclusion is not necessarily true if the material is not homogeneous or if there are temperature gradients, whether between the polarization degrees of freedom or between the baths. In fact, in the next Sections we will show that the conclusion could be different if, on the one hand, we consider non-dissipative (constant permittivity) media or, in the other hand, there are regions where the field fluctuates freely, i. e., regions where there is no material ($g(\mathbf{x})=0$) and the field is subjected to the presence of boundaries.

\subsection{Constant Dielectric Permittivity Limit}\label{CDPL}

In previous Sections we have analyzed two situations (a field in $0+1$ dimensions and a field in $n+1$ dimensions in the presence if an infinite material) where we have shown that, beyond the transient time evolution of the system, the steady regime is described only by the fluctuations of the thermal baths which are in contact with the polarization degrees of freedom of the material, as it is expected from a formalism only based on the fluctuation-dissipation theorem. This result can be seen from the final temperature dependence of the Hadamard propagator, which in the analyzed cases, was $\beta_{B}$. On the other hand, we have shown that the kernels $\mathcal{A}$ and $\mathcal{B}$, associated to the proper contribution of the field, and the contribution from the polarization degrees of freedom vanish at the steady situation (due to the dissipative dynamics of the field in every point of the space and of the polarization degrees of freedom as Brownian particles).

It is clear that these conclusions are due, physically, to the dissipative dynamics of the field in contact to reservoirs conforming the real material, which generates the damping and the absorption dominating the steady situation through its fluctuations.

We will assume now that the material is a non-dissipative dielectric, i. e., a constant permittivity material which presents no absorption and it is no dispersive because the permittivity function in the complex frequency domain is real and it is not a smooth function over the imaginary frequency axis. It is worth noting that this verifies Kramers-Kronig relations  for the complex permittivity function in the frequency domain although the function is real. In fact, Kramers-Kronig relations are not satisfied by dispersive and real permittivity functions in the imaginary frequency axis. Therefore, our calculations must include this scenario as a limiting case.

As a first step, if we clearly turn off the dissipation provided by the baths in each point of the material, we have to set $D_{\rm QBM}\equiv 0$. From the fluctuation-dissipation theorem is straightforward that $N_{\rm QBM}\equiv 0$. Therefore, this directly implies that the noise kernel also vanishes, i. e., $\mathcal{N}_{B,\mathbf{x}}\equiv 0$. This way, the bath contribution is erased from the result.

However, this is not enough because it leaves a material formed by harmonic oscillators without damping, i. e., which do not relax to a steady situation. This can be seen from the Laplace transform of the retarded Green function of the polarization degrees of freedom, which, through Eq. (\ref{RetGreenFunctionQBM}), turns out to be $\widetilde{G}_{\rm Ret,\mathbf{x}}(z)=1/(z^{2}+\omega_{\mathbf{x}}^{2})$, which presents purely imaginary poles at $z=\pm i\omega_{\mathbf{x}}$, so the retarded Green function in the time domain will be sinusoidal functions. This causes, in principle, that the contribution coming from the polarization degrees of freedom do not vanishes, i. e., $\mathcal{N}_{r,\mathbf{x}}$ not necesarily vanishes.

Nevertheless, since the dissipation kernel is $\mathcal{D}_{\mathbf{x}}=\frac{\lambda_{0,\mathbf{x}}^{2}}{2}~G_{\rm Ret,\mathbf{x}}$, and the generalized dissipation kernel  $\partial_{tt}^{2\alpha}\left[\mathcal{D}_{\mathbf{x}}(t-\tau)\right]$ that acts over the field and forms the dielectric function through its Laplace transform, will give a dispersive and real permittivity function for purely imaginary frequencies. Thus, it does not verify the Kramers-Kronig relations. Then, vanishing bath dissipation is not enough to achieve the constant dielectric limit and, in fact, it is a non-physical model.

To get a clue about how this limit can be taken, we can use the equation of motion for the field,  for arbitrary shapes of material boundaries, which in both coupling
models can be written as

\begin{eqnarray}
\partial_{\mu}\partial^{\mu}\phi+4\pi\eta_{\mathbf{x}}\lambda_{0,\mathbf{x}}^{2}~g(\mathbf{x})~\alpha~\phi-(-1)^{\alpha}8\pi\eta_{\mathbf{x}}~g(\mathbf{x})\int_{t_{0}}^{t}d\tau~\partial_{tt}^{2\alpha}\left[\mathcal{D}_{\mathbf{x}}(t-\tau)\right]~\phi(\mathbf{x},\tau)&=&0.
\label{EqMotionPhiGeneralized}
\end{eqnarray}

Going to the complex frequency domain, we Laplace-transform the associated equation for the Green function, imposing the same initial conditions as in the last Section, easily obtaining for each model

\begin{eqnarray}
\nabla^{2}\widetilde{\mathcal{G}}_{\rm Ret}-z^{2}\left[1-(-1)^{\alpha}~4\pi\eta_{\mathbf{x}}\lambda_{0,\mathbf{x}}^{2}~g(\mathbf{x})~\frac{z^{2(\alpha-1)}}{(z^{2}+\omega_{\mathbf{x}}^{2})}\right]\widetilde{\mathcal{G}}_{\rm Ret}=\delta(\mathbf{x}-\mathbf{x}').
\label{EqGreenLaplaceGeneralizedNoDiss}
\end{eqnarray}

If we now consider the equation of motion of retarded Green function, corresponding to a field subjected to the same initial conditions and with boundaries of constant dielectric permittivity $\epsilon(\mathbf{x})$ from the very beginning, we would have obtained

\begin{equation}
\nabla^{2}\widetilde{\mathcal{G}}_{\rm Ret}-z^{2}~\epsilon(\mathbf{x})~\widetilde{\mathcal{G}}_{\rm Ret}=\delta(\mathbf{x}-\mathbf{x}').
\end{equation}
which is analogous to what is found from a steady canonical quantization scheme of a field with constant permittivity dielectric boundaries \cite{Dorota1990}.

Comparing the equations it is clear that in our case we must achieve that the permittivity function given by the expression in brackets should not depend on $z$, i. e., we have to replace it by a constant. So, we can try by replacing it by its zeroth order. On the one hand, this is not possible in a simple way for the bilinear model ($\alpha=0$) because it diverges for $z=0$. On the other hand, the current-type model ($\alpha=1$) gives a finite zeroth order, allowing us to find a feasible replacement, obtaining:

\begin{eqnarray}
\nabla^{2}\widetilde{\mathcal{G}}_{\rm Ret}-z^{2}\left[1+\frac{4\pi\eta_{\mathbf{x}}\lambda_{0,\mathbf{x}}^{2}}{\omega_{\mathbf{x}}^{2}}~g(\mathbf{x})\right]\widetilde{\mathcal{G}}_{\rm Ret}=\delta(\mathbf{x}-\mathbf{x}'),
\end{eqnarray}
where it is clear that the permittivity function results $\epsilon(\mathbf{x})\equiv 1+\frac{4\pi\eta_{\mathbf{x}}\lambda_{0,\mathbf{x}}^{2}}{\omega_{\mathbf{x}}^{2}}~g(\mathbf{x})$, which correctly satisfies Kramers-Kronig relations and is constant in time.

In fact, with this replacement from the very beginning, we are removing all the dynamics of the polarization degrees of freedom and setting then equal to the steady situation in the scenario including dissipation by the evaluation at $z=0$. We clearly have that $G_{\rm Ret}\equiv 0$. All in all, it gives that the terms corresponding to the contribution of the material (polarization degrees of freedom and baths) vanishes since $\mathcal{N}\equiv 0$.

Therefore, the Hadamard propagator in Eq.(\ref{HadamardPropagator}) is up to the kernels $\mathcal{A}$ and $\mathcal{B}$, which in this case, clearly do not vanish. In fact, as long as there is Casimir force between constant dielectric boundaries due to the modification of the vacuum modes, these kernels should not vanish at the steady situation. This is in complete agreement with many results, that can be found for non-dissipative media boundaries, obtained from a steady canonical quantization scheme (see for example Ref. \cite{Dorota1990}), where quantization is carried out only by considering a Hilbert space associated to the field, and developing the Heisenberg's canonical operator method in terms of creation and annihilation field mode operators. Thus, we can write that the long-time limit ($t_{0}\rightarrow -\infty$) is given by

\begin{equation}
\mathcal{G}_{\rm H}(x_{1},x_{2})\longrightarrow 2\Big(\mathcal{A}(x_{1},x_{2})+\mathcal{B}(x_{1},x_{2})\Big).
\end{equation}

It is worth noting that, as it follows from steady canonical quantization schemes, the field's state must be taken as a thermal one, being this an additional requirement of consistency, which results in the correct thermal global factors for the correlation and Green functions in the steady situation. However, our approach naturally gives the correct thermal dependence at least when an initial high-temperature state is considered for the field, which is in agreement with the high-temperature limit of the canonical quantization or in-out formalism schemes \cite{LombiMazziRL}.

Finally, we have shown a first and simplest example where the kernels $\mathcal{A}$ and $\mathcal{B}$ do not vanish at the steady situation and in fact, in this case, they define the long-time regime. However, this is not totally new because we clearly know that there exist Casimir force between constant dielectric boundaries due to the modified vacuum modes. Anyway, we have just proved that our approach correctly reproduce that situation as a limiting case. In the next Section, we will study another situation where these terms do not vanish but neither define completely the long-time regime.

\subsection{Field and Material Boundaries}\label{FAMB}

In this Section, let us study a particular situation of the presence of boundaries. At this point, we have already seen that for the case of a $0+1$ dimensions field of frequency $\Omega$, the long-time regime is defined by the bath's contribution, while the polarization degree of freedom and the proper field contributions vanish at the steady situation. Then, we have also seen that a $n+1$ dimensions scalar field, interacting with homogeneous material all over the space, can be reduced to an infinite set of $0+1$ fields with frequency $k$, representing the field modes that evolve in time due to the sudden appearance of the material. We have shown that in the long-time limit, as in the $0+1$ case, the only contribution to the energy-momentum tensor that survives is also the one associated to the baths. The polarization degrees of freedom and the own field have vanishing contributions at the steady situation.

However, although we were tempted to assume that the result of the last two Sections is quite general and always valid, we have presented a limiting case where the reverse is true and the annulation of the dissipation makes that the kernels $\mathcal{A}$ and $\mathcal{B}$ become the responsible of the Casimir force between non-dissipative boundaries in the long-time regime.

As we pointed out before, this is not the only case where these kernels contribute to the steady situation. If the material is inhomogeneous or there exist regions without material
(i.e., vacuum regions that define material boundaries) the same could be true. Therefore, this Section gives a simple example of the presence of boundaries and the analysis of the steady situation.

\subsubsection{The Retarded Green Function}

Back to the field equation for general boundaries given in Eq.(\ref{EqMotionPhiGeneralized}) for both coupling models, we can again Laplace transform the associated equation for the retarded Green function with appropriate initial conditions to obtain

\begin{eqnarray}
\nabla^{2}\widetilde{\mathcal{G}}_{\rm Ret}-z^{2}\left[1-(-1)^{\alpha}~4\pi\eta_{\mathbf{x}}\lambda_{0,\mathbf{x}}^{2}~g(\mathbf{x})~z^{2(\alpha-1)}~\widetilde{G}_{\rm{Ret},\mathbf{x}}(z)\right]\widetilde{\mathcal{G}}_{\rm Ret}=\delta(\mathbf{x}-\mathbf{x}'),
\label{EqGreenLaplaceGeneralized}
\end{eqnarray}
where $\widetilde{\mathcal{G}}_{\rm Ret}$ results to be the inverse of the differential operator $\nabla^{2}-z^{2}\left[1-(-1)^{\alpha}~4\pi\eta_{\mathbf{x}}\lambda_{0,\mathbf{x}}^{2}~g(\mathbf{x})~z^{2(\alpha-1)}~\widetilde{G}_{\rm{Ret},\mathbf{x}}(z)\right]$, i.e., it is directly the Green function associated to this operator.

We will consider a single homogeneous Dirac delta plate located at $x_{\perp}=0$ ($x_{\perp},\mathbf{x}_{\parallel}$ are the orthogonal and parallel coordinates to the plate of a given space point $\mathbf{x}$), which is described by the material distribution $g(\mathbf{x})\equiv\delta(x_{\perp})$. Thus, Eq.(\ref{EqGreenLaplaceGeneralized}) results

\begin{eqnarray}
\nabla^{2}\widetilde{\mathcal{G}}_{\rm Ret}-z^{2}\left[1-(-1)^{\alpha}~4\pi\eta\lambda_{0}^{2}~\delta(x_{\perp})~z^{2(\alpha-1)}~\widetilde{G}_{\rm{Ret}}(z)\right]\widetilde{\mathcal{G}}_{\rm Ret}=\delta(\mathbf{x}-\mathbf{x}').
\label{EqGreenLaplaceGeneralizedDiracHomogenea}
\end{eqnarray}

It is clear that the last equation presents translational invariance on the parallel coordinates $\mathbf{x}_{\parallel}$, so the Green function must depends on $\mathbf{x}_{\parallel}-\mathbf{x}'_{\parallel}$. Then,

\begin{eqnarray}
\frac{\partial^{2}\widetilde{\mathcal{G}}_{\rm Ret}}{\partial x_{\perp}^{2}}-(z^{2}+k_{\parallel}^{2})~\widetilde{\mathcal{G}}_{\rm Ret}+(-1)^{\alpha}~4\pi\eta\lambda_{0}^{2}~\delta(x_{\perp})~z^{2\alpha}~\widetilde{G}_{\rm{Ret}}(z)~\widetilde{\mathcal{G}}_{\rm Ret}=\delta(x_{\perp}-x'_{\perp}).
\label{EqGreenLaplaceGeneralizedDiracHomogeneaFourierParallel}
\end{eqnarray}
where $k_{\parallel}=|\mathbf{k}_{\parallel}|$ and $\widetilde{\mathcal{G}}_{\rm Ret}\equiv\widetilde{\mathcal{G}}_{\rm Ret}(x_{\perp},x'_{\perp},k_{\parallel},z)$.

It is worth noting that the last equation turns out to be a Sturn-Liouville equation for the Green function, so it can be calculated by the technique described in Ref. \cite{Collin}, where it is constructed as

\begin{equation}
\widetilde{\mathcal{G}}_{\rm Ret}(x_{\perp},x'_{\perp},k_{\parallel},z)=\frac{\Phi^{(L)}(x_{<})~\Phi^{(R)}(x_{>})}{W(x'_{\perp})},
\label{GreenFunctionCollin}
\end{equation}
where $x_{<}$ ($x_{>}$) is the smaller (bigger) between $x_{\perp}$ and $x'_{\perp}$, $W(x)=\Phi^{(L)}(x)~\frac{d\Phi^{(R)}}{dx}-\frac{d\Phi^{(L)}}{dx}~\Phi^{(R)}(x)$ is the Wronskian (which has to be a constant function) of the solutions $\{\Phi^{(L)},\Phi^{(R)}\}$, which are two homogeneous solutions $\Phi^{(L,R)}$ that satisfy the associated homogeneous equation

\begin{eqnarray}
\frac{\partial^{2}\Phi}{\partial x_{\perp}^{2}}-(z^{2}+k_{\parallel}^{2})~\Phi+(-1)^{\alpha}~4\pi\eta\lambda_{0}^{2}~\delta(x_{\perp})~z^{2\alpha}~\widetilde{G}_{\rm{Ret}}(z)~\Phi=0.
\label{EqSolutionsPhi}
\end{eqnarray}
and the boundary condition in one of the two range endpoints, i.e., $\Phi^{L}$ ($\Phi^{R}$) satisfies the boundary condition in the left (right) endpoint of the interval. In our case, that boundary condition is to have outgoing waves in the corresponding region including the respective endpoint.

The presence of a Dirac delta function in one of the terms of the equation makes that we will obtain the solution in two regions, each one with positive and negative coordinates $x_{\perp}$ respectively; and on the other hand, it gives a boundary condition with a jolt on the derivative, which can be obtained from the equation itself by integrating over an interval containing the root of the delta function and then take its length to zero around the root, clearly obtaining

\begin{equation}
\frac{\partial\Phi}{\partial x_{\perp}}\Big|_{x_{\perp}=0^{+}}-\frac{\partial\Phi}{\partial x_{\perp}}\Big|_{x_{\perp}=0^{-}}=(-1)^{\alpha}~4\pi\eta\lambda_{0}^{2}~z^{2\alpha}~\widetilde{G}_{\rm Ret}(z)~\Phi(0),
\label{BoundaryConditionDerDirac}
\end{equation}
which goes together with the continuity of the solution.

Therefore, in each region, the solutions are plane waves so, after imposing the boundary conditions, both solutions result

\begin{eqnarray}
\Phi^{(L)}(x_{\perp})= \left\{
\begin{array}{lr rl}
t~e^{\sqrt{z^{2}+k_{\parallel}^{2}}~x_{\perp}}, &&& \text{for}~x_{\perp}<0\\
e^{\sqrt{z^{2}+k_{\parallel}^{2}}~x_{\perp}}+r~e^{-\sqrt{z^{2}+k_{\parallel}^{2}}~x_{\perp}},  &&&\text{for}~0<x_{\perp}\\
\end{array}
\right.
\label{PhiSolutionL}
\end{eqnarray}

\begin{eqnarray}
\Phi^{(R)}(x_{\perp})= \left\{
\begin{array}{lr rl}
e^{-\sqrt{z^{2}+k_{\parallel}^{2}}~x_{\perp}}+r~e^{\sqrt{z^{2}+k_{\parallel}^{2}}~x_{\perp}}, &&& \text{for}~x_{\perp}<0\\
t~e^{-\sqrt{z^{2}+k_{\parallel}^{2}}~x_{\perp}},  &&&\text{for}~0<x_{\perp}\\
\end{array}
\right.
\label{PhiSolutionR}
\end{eqnarray}
where $r$ and $t$ are the reflection and transmission coefficients for one plate respectively and given by:

\begin{equation}
r=-(-1)^{\alpha}~2\pi\eta\lambda_{0}^{2}~\frac{z^{2\alpha}}{\sqrt{z^{2}+k_{\parallel}^{2}}}~\widetilde{G}_{\rm Ret}(z)~t~~~~~~~,~~~~~~~t=\frac{1}{\left(1+(-1)^{\alpha}~2\pi\eta\lambda_{0}^{2}~\frac{z^{2\alpha}}{\sqrt{z^{2}+k_{\parallel}^{2}}}~\widetilde{G}_{\rm Ret}(z)\right)},
\label{CoefficientsRTnD}
\end{equation}
where it is clear that $t=1+r$.

Then, the Laplace-Fourier-transform of the retarded Green function for a field point $x_{\perp}<0$ follows:

\begin{eqnarray}
\widetilde{\mathcal{G}}_{\rm Ret}(x_{\perp},x'_{\perp},k_{\parallel},z)=-\frac{1}{2\sqrt{z^{2}+k_{\parallel}^{2}}}\left\{
\begin{array}{lr rl}
e^{\sqrt{z^{2}+k_{\parallel}^{2}}~x'_{\perp}}\left(e^{-\sqrt{z^{2}+k_{\parallel}^{2}}~x_{\perp}}+r~e^{\sqrt{z^{2}+k_{\parallel}^{2}}~x_{\perp}}\right), &&& \text{for}~x'_{\perp}<x_{\perp}<0\\
\left(e^{-\sqrt{z^{2}+k_{\parallel}^{2}}~x'_{\perp}}+r~e^{\sqrt{z^{2}+k_{\parallel}^{2}}~x'_{\perp}}\right)e^{\sqrt{z^{2}+k_{\parallel}^{2}}~x_{\perp}},  &&&\text{for}~x_{\perp}<x'_{\perp}<0\\
t~e^{\sqrt{z^{2}+k_{\parallel}^{2}}~(x_{\perp}-x'_{\perp})}.  &&&\text{for}~x_{\perp}<0<x'_{\perp}\\
\end{array}
\right.
\label{LaplaceFourierGreenFunctionDirac}
\end{eqnarray}

For simplicity in the calculations, we continue with the one-dimensional version of the problem, i. e., the case of a  $1+1$ field where the only dimension of interest clearly is the one associated to the perpendicular coordinate $x_{\perp}$, which we call now $x$. Therefore, to obtain the results for this case we also have to discard everything related to the parallel dimensions. We can do this simply by setting $k_{\parallel}$ equal to $0$ in all the results. This simplifies all the expressions and the Laplace transform of the retarded Green function in Eq.(\ref{LaplaceFourierGreenFunctionDirac}) is

\begin{eqnarray}
\widetilde{\mathcal{G}}_{\rm Ret}(x,x',z)=-\frac{1}{2z}\left\{
\begin{array}{lr rl}
e^{zx'}\left(e^{-zx}+r~e^{zx}\right), &&& \text{for}~x'<x<0\\
\left(e^{-zx'}+r~e^{zx'}\right)e^{zx},  &&&\text{for}~x<x'<0\\
t~e^{z(x-x')},  &&&\text{for}~x<0<x'\\
\end{array}
\right.
\label{LaplaceFourierGreenFunctionDirac1D}
\end{eqnarray}
where the reflection and transmission coefficients are now given by:

\begin{equation}
r=-(-1)^{\alpha}~2\pi\eta\lambda_{0}^{2}~z^{2\alpha-1}~\widetilde{G}_{\rm Ret}(z)~t~~~~~~~,~~~~~~~t=\frac{1}{\left(1+(-1)^{\alpha}~2\pi\eta\lambda_{0}^{2}~z^{2\alpha-1}~\widetilde{G}_{\rm Ret}(z)\right)}.
\label{CoefficientsRT1D}
\end{equation}

We can transform Laplace back, by Mellin's formula and the Residue theorem \cite{Schiff}, assuming that the poles of the Laplace transform of the retarded Green function have non-positive real parts. It is important to remark that, following the discussion done in Ref.\cite{LombiMazziRL}, for both coupling models, this can always be ensured by introducing an appropriate cut-off function in the Laplace transform of the dissipation kernel $\widetilde{D}_{\rm QBM}$ (in fact, any spectral density that characterizes the environment has a physical cutoff function). Moreover, assuming that the dissipation (represented by $D_{\rm QBM}$), the frequency $\omega$ and coupling constant $\lambda_{0}$ are not zero, the only pole with vanishing real part is the one at $z=0$, which appears (for each coupling case) in different terms of the Laplace transform of retarded Green function, resulting in different behaviors of the retarded Green function. This can be seen working out the last expression of the reflection coefficient $r$ in each model. However, this pole do not change the conclusions of the present Section, so we will continue the analysis without losing generality.

Therefore, the Green function can be formally written as:

\begin{eqnarray}
\mathcal{G}_{\rm Ret}(x,x',t)=-\frac{1}{2}\left\{
\begin{array}{lr rl}
\Theta(x'-x+t)+\Theta(x+x'+t)\left[\alpha-1+\sum_{z_{j}}R_{j}~e^{z_{j}(x+x'+t)}\right], &&& \text{for}~x'<x<0\\
\Theta(x-x'+t)+\Theta(x+x'+t)\left[\alpha-1+\sum_{z_{j}}R_{j}~e^{z_{j}(x+x'+t)}\right],  &&&\text{for}~x<x'<0\\
\Theta(x-x'+t)\left[\alpha+\sum_{z_{j}}R_{j}~e^{z_{j}(x-x'+t)}\right],  &&&\text{for}~x<0<x'\\
\end{array}
\right.
\label{GreenFunctionDirac1DDisplay}
\end{eqnarray}
where $z_{j}$ are all the poles of $r$ with negative real part, i.e., the pole at $z=0$ is calculated explicitly in each model.  For the others poles we have $R_{j}\equiv Res\left[\frac{r}{z},z_{j}\right]=Res\left[\frac{t}{z},z_{j}\right]$. Given that the retarded Green function must be real, its poles must come in pairs (i.e., if $z_{j}$ is a pole then its conjugate $z_{j}^{*}$ is a pole too) unless $z_{j}$ is real.

This expression, however, can be worked out by re-arranging the terms and combining their Heavyside functions to obtain a suitable closed form for the
retarded Green function in each model for a field point $x<0$

\begin{eqnarray}
\mathcal{G}_{\rm Ret}(x,x',t)&=&\mathcal{G}_{\rm Ret}^{0}(x,x',t)+\frac{(1-\alpha)}{2}~\Theta(-x)~\Theta(x+x'+t)~\Theta(x-x'+t)\nonumber\\
&&-\frac{\Theta(-x)}{2}\sum_{z_{j}}R_{j}~e^{z_{j}(x+t)}\left[e^{z_{j}x'}~\Theta(-x')~\Theta(x+x'+t)+e^{-z_{j}x'}~\Theta(x')~\Theta(x-x'+t)\right],
\label{GreenFunctionDirac1DClosedForm}
\end{eqnarray}
where $\mathcal{G}_{\rm Ret}^{0}(x,x',t)\equiv-\frac{\Theta(-x)}{2}~\Theta(x'-x+t)~\Theta(x-x'+t)$ is the retarded Green function in free space for a field point $x<0$.

It is worth noting, on the one hand, that the second term is an extra term only for the bilinear model due to the presence of the plate but independent of the material properties. On the other hand, the third term is directly and entirely related to the presence of the plate and it contains all the information about the material contribution to the transient evolution (i.e., relaxation) and the new steady situation that the field will achieve. It is clear that it implicitly depends on the coupling model because the poles $z_{j}$ depend on it.

From Eq.(\ref{GreenFunctionDirac1DClosedForm}) it can be easily proved, by looking carefully the products of distributions, that the time derivative of the retarded Green function has a simple form given by:

\begin{eqnarray}
\dot{\mathcal{G}}_{\rm Ret}(x,x',t)&=&\dot{\mathcal{G}}_{\rm Ret}^{0}(x,x',t)\nonumber \\
&-&\frac{\Theta(-x)}{2}\sum_{z_{j}}z_{j}~R_{j}~e^{z_{j}(x+t)}\left[e^{z_{j}x'}~\Theta(-x')~\Theta(x+x'+t)+e^{-z_{j}x'}~\Theta(x')~\Theta(x-x'+t)\right],
\label{DerivativeGreenFunctionDirac1DClosedForm}
\end{eqnarray}
where in this expression the only difference between the coupling models relies on the poles $z_{j}$.

\subsubsection{The Long-Time Regime}

With the retarded Green function for the present problem, we can proceed to study some dynamical aspects and features about the steady situation.

As we just obtained in previous sections, our interest is the Hadamard propagator given in Eq.(\ref{HadamardPropagator}), which we can use to calculate the expectation value of the energy-momentum components through Eq.(\ref{TmunuExpValue}). As is stated by Eq.(\ref{HadamardPropagator}), the Hadamard propagator has several contributions that can be divided in two parts, one coming from the field generated by all the components of the material (polarization degrees of freedom and baths) which is represented by the noise kernel $\mathcal{N}$, and another one coming from the field generated by the vacuum fluctuations subjected to the actual boundary conditions, which is represented by the kernels $\mathcal{A}$ and $\mathcal{B}$ and will imply a modification of the field modes through a transient evolution from the initial free field to the new steady field.

Let's study firstly the material contribution. As we have proved in Sec.\ref{FIIM}, when the material is modeled as Brownian particles interacting with the field by both coupling models, the material contribution at the steady situation have only the contributions coming from the baths, while the particles merely act as a bridge connecting the field with the baths, but having no contribution in the long-time regime due to their dissipative Brownian dynamics. This was basically contained in the fact that in the long-time limit ($t_{0}\rightarrow-\infty$), we clearly have that $\mathcal{N}\rightarrow\mathcal{N}_{B}$. In the present case, although there are regions without material, the result is still valid. It is clear that in this case the Green function is given by Eq.(\ref{GreenFunctionDirac1DClosedForm}) but the formal expression is the same. In fact, it is worth noting also that the material distribution $g$ will define the range of integration, having no contribution from the points outside the material.

On the other hand, we have the contribution to the field generated by the vacuum fluctuations represented by $\mathcal{A}$ and $\mathcal{B}$. We are tempted to assume that, as in Sec.\ref{FIIM}, these contributions vanish at the steady situation giving only, a transient behavior. However, as we just pointed out at the end of that Section, this could not be true when there are vacuum regions where the field fluctuates freely.

Therefore, considering the kernel $\mathcal{A}$ in Eq.(\ref{KernelAHighT}) and the expression for the retarded Green function given in Eq.(\ref{GreenFunctionDirac1DClosedForm}), is clear that the product $\mathcal{G}_{\rm Ret}(x_{1},x,t_{1}-t_{0})~\mathcal{G}_{\rm Ret}(x_{2},x,t_{2}-t_{0})$ will have at most nine terms (depending which coupling model we are considering) due to all the possible combinations of the three separated terms which then have to be integrated over $x$.  Using the symmetry of the kernel, the number of integrals to calculate is six at most. The complication in the full calculation is due to the fact that each integral involves products of distributions having the integration variable and both field points $(x_{1},t_{1})$ and $(x_{2},t_{2})$; results will depend on multiple relations between the coordinates of the field points.

On the other hand, the full calculation of kernel $\mathcal{B}$ is so complicated as for the previous kernel. In the one dimensional case, it is easy to calculate the kernel $K$ in Eq.(\ref{KernelKHighTCoordinate}) through the Residue theorem, obtaining that $K(x-x')=-\frac{|x-x'|}{2\beta_{\phi}}$, which can be written as two terms. Then, kernel $\mathcal{B}$ involves a double integration (over $x$ and $x'$) of the triple product $\dot{\mathcal{G}}_{\rm Ret}(x_{1},x,t_{1}-t_{0})~K(x-x')~\dot{\mathcal{G}}_{\rm Ret}(x_{2},x',t_{2}-t_{0})$. From Eq.(\ref{DerivativeGreenFunctionDirac1DClosedForm}) it is clear that the derivative of the retarded Green function has two terms, so to obtain $\mathcal{B}$ the number of integrals to perform is in principle eight. Due to the symmetry, the final number of double integrals reduce to six for this kernel too.

As we are interested in general features about the transient time evolution and the steady situation, we will not proceed to a complete calculation of the terms, but we will show that there are steady terms associated to the contribution of these kernels.

We should note that the terms in the kernels $\mathcal{A}$ and $\mathcal{B}$ associated to the products of the free field retarded Green function $\mathcal{G}_{\rm Ret}^{0}$ and its derivative, i.e., the terms $\mathcal{G}_{\rm Ret}^{0}(x_{1},x,t_{1}-t_{0})~\mathcal{G}_{\rm Ret}^{0}(x_{2},x,t_{2}-t_{0})$ in $\mathcal{A}$ and $\dot{\mathcal{G}}_{\rm Ret}^{0}(x_{1},x,t_{1}-t_{0})~K(x-x')~\dot{\mathcal{G}}_{\rm Ret}^{0}(x_{2},x',t_{2}-t_{0})$ in $\mathcal{B}$ are the ones that will be removed by the Casimir prescription (subtraction with the free field case), so we do not have to calculate them.

We should note also that the crossed terms (i.e. terms combining different terms of the Green function) will be transient terms since those integrations will generate constant terms that will vanish in the derivatives and through the limit needed to calculate the expectation values of the energy-momentum tensor, or terms that will exponentially decay at the long-time limit, or divergent terms that must be subtracted to define a correct (non-divergent) Hadamard propagator. As we are interested now in the steady situation, we will not calculate them.

Independently of which model we are considering, to study the long-time regime, the products involving two sums over poles will be the ones that result in steady contributions. As a first case, we consider the corresponding term found in the kernel $\mathcal{A}$ for field point $x_{1,2}<0$:

\begin{eqnarray}
\mathcal{A}(x_{1},x_{2},t_{1},t_{2})&=&(\text{Free Field Terms})+(\text{Crossed Terms})+\frac{\Theta(-x_{1})~\Theta(-x_{2})}{4\beta_{\phi}}\sum_{j,l}R_{j}~R_{l}~e^{z_{j}(x_{1}+t_{1}-t_{0})}~e^{z_{l}(x_{2}+t_{2}-t_{0})}\nonumber\\
&&\times\int_{-\infty}^{+\infty}dx\left[e^{z_{j}x}~\Theta(-x)~\Theta(x_{1}+x+t_{1}-t_{0})+e^{-z_{j}x}~\Theta(x)~\Theta(x_{1}-x+t_{1}-t_{0})\right]\nonumber\\
&&~~~~~~~~~~~\times\left[e^{z_{l}x}~\Theta(-x)~\Theta(x_{2}+x+t_{2}-t_{0})+e^{-z_{l}x}~\Theta(x)~\Theta(x_{2}-x+t_{2}-t_{0})\right].
\label{KernelAHighTDirac1DSteadyTerms}
\end{eqnarray}

Considering that $\Theta(x)~\Theta(-x)\equiv 0$ and $\Theta(\pm x)~\Theta(\pm x)\equiv\Theta(\pm x)$, there are vanishing integrals in the expression. Then, making a substitution $x\rightarrow -x$ on one of the two resulting terms, all the integrals show to be the same, clearly obtaining:

\begin{eqnarray}
\mathcal{A}(x_{1},x_{2},t_{1},t_{2})&=&(\text{Free Field Terms})+(\text{Crossed Terms})\nonumber\\
&+&\frac{\Theta(-x_{1})\Theta(-x_{2})}{2\beta_{\phi}}\sum_{j,l}R_{j}~R_{l}~e^{z_{j}(x_{1}+t_{1}-t_{0})}~e^{z_{l}(x_{2}+t_{2}-t_{0})}\nonumber \\
&\times & \int_{-\infty}^{+\infty}dx~e^{(z_{j}+z_{l})x}~\Theta(-x)~\Theta(x_{1}+x+t_{1}-t_{0})~\Theta(x_{2}+x+t_{2}-t_{0}).
\end{eqnarray}

Considering that $\Theta(x_{1}+x+t_{1}-t_{0})~\Theta(x_{2}+x+t_{2}-t_{0})=\Theta(x_{1}-x_{2}+t_{1}-t_{2})~\Theta(x_{2}+x+t_{2}-t_{0})+\Theta(x_{2}-x_{1}+t_{2}-t_{1})~\Theta(x_{1}+x+t_{1}-t_{0})$, the last integral can be easily calculated

\begin{eqnarray}
&&\mathcal{A}(x_{1},x_{2},t_{1},t_{2})=(\text{Free Field Terms})+(\text{Crossed Terms})+\frac{\Theta(-x_{1})~\Theta(-x_{2})}{2\beta_{\phi}}\sum_{j,l}\frac{R_{j}~R_{l}}{(z_{j}+z_{l})}\nonumber\\
&&\times\left[\Big(\Theta(x_{1}-x_{2}+t_{1}-t_{2})~\Theta(x_{2}+t_{2}-t_{0})+\Theta(x_{2}-x_{1}+t_{2}-t_{1})~\Theta(x_{1}+t_{1}-t_{0})\Big)~e^{z_{j}(x_{1}+t_{1}-t_{0})}~e^{z_{l}(x_{2}+t_{2}-t_{0})}\right. \\
&&\left.-~\Theta(x_{1}-x_{2}+t_{1}-t_{2})~\Theta(x_{2}+t_{2}-t_{0})~e^{z_{j}(x_{1}-x_{2}+t_{1}-t_{2})}-\Theta(x_{2}-x_{1}+t_{2}-t_{1})~\Theta(x_{1}+t_{1}-t_{0})~e^{z_{l}(x_{2}-x_{1}+t_{2}-t_{1})}\right].\nonumber
\label{KernelAHighTDirac1DSteadyTermsFinal}
\end{eqnarray}

On the other hand, the kernel $\mathcal{B}$ presents a more complicated structure because it involves two integrations (one over $x$ and other one over $x'$) and an extra kernel $K(x-x')$ which couples both integrations preventing a separate calculation. Following the same train of thought, we focus in the terms involving two sums over poles. Therefore, kernel $\mathcal{B}$ reads:

\begin{eqnarray}
\mathcal{B}(x_{1},x_{2},t_{1},t_{2})&=&(\text{Free Field Terms})+(\text{Crossed Terms})\nonumber \\ &-&\frac{\Theta(-x_{1})~\Theta(-x_{2})}{8\beta_{\phi}}\sum_{j,l}z_{j}~z_{l}~R_{j}~R_{l}~e^{z_{j}(x_{1}+t_{1}-t_{0})}~e^{z_{l}(x_{2}+t_{2}-t_{0})}\nonumber\\
&&\times\int_{-\infty}^{+\infty}dx\int_{-\infty}^{+\infty}dx'~|x-x'|\left[e^{z_{j}x}~\Theta(-x)~\Theta(x_{1}+x+t_{1}-t_{0})+e^{-z_{j}x}~\Theta(x)~\Theta(x_{1}-x+t_{1}-t_{0})\right]\nonumber\\
&& \times\left[e^{z_{l}x'}~\Theta(-x')~\Theta(x_{2}+x'+t_{2}-t_{0})+e^{-z_{l}x'}~\Theta(x')~\Theta(x_{2}-x'+t_{2}-t_{0})\right].
\end{eqnarray}

The integration over $x'$ can be done first by writing $|x-x'|=\Theta(x-x')~(x-x')+\Theta(x'-x)~(x'-x)$. Working out the integral, we obtain that the result can be separated again in terms that will be part of the transient evolution and will vanish at the long-time regime and terms that will give steady results. In fact, the integral can be written as:

\begin{eqnarray}
\int_{-\infty}^{+\infty}&dx'&~|x-x'|\left[e^{z_{l}x'}~\Theta(-x')~\Theta(x_{2}+x'+t_{2}-t_{0})+e^{-z_{l}x'}~\Theta(x')~\Theta(x_{2}-x'+t_{2}-t_{0})\right]=\nonumber\\
&=&\frac{2}{z_{l}^{2}}\left[\Theta(-x)~\Theta(x_{2}+x+t_{2}-t_{0})~e^{z_{l}x}+\Theta(x)~\Theta(x_{2}-x+t_{2}-t_{0})~e^{-z_{l}x}\right]+(\text{Transient Terms}).
\end{eqnarray}

Thus, kernel $\mathcal{B}$ reads:

\begin{eqnarray}
&&\mathcal{B}(x_{1},x_{2},t_{1},t_{2})=(\text{Free Field Terms})+(\text{Crossed Terms})+(\text{Transient Terms})  \\ &-&\frac{\Theta(-x_{1})~\Theta(-x_{2})}{4\beta_{\phi}}\sum_{j,l}\frac{z_{j}}{z_{l}}~R_{j}~R_{l}\nonumber\\
&&\times~e^{z_{j}(x_{1}+t_{1}-t_{0})}~e^{z_{l}(x_{2}+t_{2}-t_{0})}\int_{-\infty}^{+\infty}dx\left[e^{z_{j}x}~\Theta(-x)~\Theta(x_{1}+x+t_{1}-t_{0})+e^{-z_{j}x}~\Theta(x)~\Theta(x_{1}-x+t_{1}-t_{0})\right]\nonumber\\
&&\times\left[\Theta(-x)~\Theta(x_{2}+x+t_{2}-t_{0})~e^{z_{l}x}+\Theta(x)~\Theta(x_{2}-x+t_{2}-t_{0})~e^{-z_{l}x}\right],\nonumber
\end{eqnarray}
where it is worth noting that the resulting integral is the same as the one in kernel $\mathcal{A}$ in Eq.(\ref{KernelAHighTDirac1DSteadyTerms}).

Then, the result is the same and the kernel can be written as:

\begin{eqnarray}
&&\mathcal{B}(x_{1},x_{2},t_{1},t_{2})=(\text{Free Field Terms})+(\text{Crossed Terms})+(\text{Transient Terms}) \\ &-& \frac{\Theta(-x_{1})~\Theta(-x_{2})}{2\beta_{\phi}}\sum_{j,l}\frac{z_{j}}{z_{l}}~\frac{R_{j}~R_{l}}{(z_{j}+z_{l})}\nonumber\\
&&\times\left[\Big(\Theta(x_{1}-x_{2}+t_{1}-t_{2})~\Theta(x_{2}+t_{2}-t_{0})+\Theta(x_{2}-x_{1}+t_{2}-t_{1})~\Theta(x_{1}+t_{1}-t_{0})\Big)~e^{z_{j}(x_{1}+t_{1}-t_{0})}~e^{z_{l}(x_{2}+t_{2}-t_{0})}\right.\nonumber\\
&&\left.-~\Theta(x_{1}-x_{2}+t_{1}-t_{2})~\Theta(x_{2}+t_{2}-t_{0})~e^{z_{j}(x_{1}-x_{2}+t_{1}-t_{2})}-\Theta(x_{2}-x_{1}+t_{2}-t_{1})~\Theta(x_{1}+t_{1}-t_{0})~e^{z_{l}(x_{2}-x_{1}+t_{2}-t_{1})}\right].\nonumber
\end{eqnarray}

By considering this last equation and Eq.(\ref{KernelAHighTDirac1DSteadyTermsFinal}), it is now straightforward that the proper field contribution, given by the sum of the kernels $\mathcal{A}$ and $\mathcal{B}$ can be written as:

\begin{eqnarray}
&&\mathcal{A}(x_{1},x_{2},t_{1},t_{2})+\mathcal{B}(x_{1},x_{2},t_{1},t_{2})=(\text{Free Field Terms})+(\text{Crossed Terms})+(\text{Transient Terms}) \\
&+&\frac{\Theta(-x_{1})~\Theta(-x_{2})}{2\beta_{\phi}}\sum_{j,l}\left(1-\frac{z_{j}}{z_{l}}\right)~\frac{R_{j}~R_{l}}{(z_{j}+z_{l})}\nonumber\\
&&\times\left[\Big(\Theta(x_{1}-x_{2}+t_{1}-t_{2})~\Theta(x_{2}+t_{2}-t_{0})+\Theta(x_{2}-x_{1}+t_{2}-t_{1})~\Theta(x_{1}+t_{1}-t_{0})\Big)~e^{z_{j}(x_{1}+t_{1}-t_{0})}~e^{z_{l}(x_{2}+t_{2}-t_{0})}\right.\nonumber\\
&&\left.-~\Theta(x_{1}-x_{2}+t_{1}-t_{2})~\Theta(x_{2}+t_{2}-t_{0})~e^{z_{j}(x_{1}-x_{2}+t_{1}-t_{2})}-\Theta(x_{2}-x_{1}+t_{2}-t_{1})~\Theta(x_{1}+t_{1}-t_{0})~e^{z_{l}(x_{2}-x_{1}+t_{2}-t_{1})}\right]. \nonumber
\end{eqnarray}

The last two terms in the brackets contain exponentials whose exponents do not depend on the initial time $t_{0}$. Therefore those terms will not vanish at the long-time limit ($t_{0}\rightarrow-\infty$). This shows that a part of the proper field contribution has not only transient but also steady terms that contributes to the long-time regime. Note, in fact, that these terms in the Hadamard propagator will result as constant terms in the expectation values of the energy-momentum tensor components of Eq.(\ref{TmunuExpValue}) after differentiating and calculating the coincidence limit.

Moreover, we can work out these terms and write them in a more familiar way connecting with previous works. It should be noted that in the first term (associated to $e^{z_{j}(x_{1}-x_{2}+t_{1}-t_{2})}$) the sum over $l$ can be worked out through the Residue theorem, while in the second term (associated to $e^{z_{l}(x_{2}-x_{1}+t_{2}-t_{1})}$) the sum over $j$ can be done.

Then, let us take firstly this last sum over $j$. Considering that all the poles are simple and $R_{j}\equiv Res\left[\frac{r}{z},z_{j}\right]$, we can write:

\begin{equation}
\sum_{j}\frac{(z_{l}-z_{j})}{(z_{j}+z_{l})}~R_{j}=\sum_{j}Res\left[\frac{(z_{l}-z)}{(z+z_{l})}\frac{r}{z},z_{j}\right].
\end{equation}

From Eq.(\ref{CoefficientsRT1D}) and given that $Re(z_{l})<0$ for every pole $z_{l}$ (so $z_{l}+z_{j}\neq 0$), we can show that the complex function $\frac{(z_{l}-z)}{(z+z_{l})}\frac{r}{z}$ goes to $0$ when $|z|\rightarrow+\infty$, independently of the direction in the complex plane, and that its set of poles is given by all the poles $z_{j}$, the pole $-z_{l}$ (which depend on the term on the sum over $l$ that we are considering) and the pole $0$ only in the bilinear coupling model. Therefore, through the Residue theorem, for a circle $\mathcal{C}_{R}^{+}$ of radius $R$ in the complex plane  that contains all the poles, when $R\rightarrow+\infty$, we can write:

\begin{eqnarray}
&&0=\int_{\mathcal{C}_{\infty}}\frac{dz}{2\pi i}~\frac{(z_{l}-z)}{(z+z_{l})}\frac{r}{z}=\sum_{j}Res\left[\frac{(z_{l}-z)}{(z+z_{l})}\frac{r}{z},z_{j}\right]-2~r(-z_{l})+\alpha-1,
\end{eqnarray}
where the last two terms are the results of calculating explicitly the poles at $-z_{l}$ and at $0$.

Therefore, the whole term associated to $e^{z_{l}(x_{2}-x_{1}+t_{2}-t_{1})}$ reads:

\begin{eqnarray}
\sum_{j,l}\left(1-\frac{z_{j}}{z_{l}}\right)~\frac{R_{j}~R_{l}}{(z_{j}+z_{l})}~e^{z_{l}(x_{2}-x_{1}+t_{2}-t_{1})}=\sum_{l}\frac{R_{l}}{z_{l}}\Big(2~r(-z_{l})+1-\alpha\Big)~e^{z_{l}(x_{2}-x_{1}+t_{2}-t_{1})}.
\label{SumaL}
\end{eqnarray}

Analogously, we can proceed for the other term, associated to $e^{z_{j}(x_{1}-x_{2}+t_{1}-t_{2})}$, by starting with the sum over $l$. The calculation is the same but over the complex function $\frac{(z-z_{j})}{(z+z_{j})}\frac{r}{z^{2}}$ and except that the pole at $z=0$ is simple for the current-type model, while is of second order in the bilinear model. Then we finally have:

\begin{eqnarray}
\sum_{j,l}\left(1-\frac{z_{j}}{z_{l}}\right)~\frac{R_{j}~R_{l}}{(z_{j}+z_{l})}~e^{z_{j}(x_{1}-x_{2}+t_{1}-t_{2})}=\sum_{j}R_{j}\Bigg(2~\frac{r(-z_{j})}{z_{j}}+\frac{2\pi\eta\lambda_{0}^{2}}{\omega^{2}}~\alpha+(1-\alpha)\frac{\omega^{2}}{2\pi\eta\lambda_{0}^{2}}\Bigg)~e^{z_{j}(x_{1}-x_{2}+t_{1}-t_{2})}
\label{SumaJ}
\end{eqnarray}
where the difference in units between the last two terms in the brackets is due to the fact that the coupling constant $\lambda_{0}$ change its units depending on the coupling model.

The last terms involves differences between both coupling models in Eqs.(\ref{SumaL}) and (\ref{SumaJ}). It can be shown that are divergent terms in the coincidence limit, so we discard them by regularizing the expression. This way, we can write the proper field contribution as:

\begin{eqnarray}
\mathcal{A}(x_{1},x_{2},t_{1},t_{2})&+&\mathcal{B}(x_{1},x_{2},t_{1},t_{2})=(\text{Free Field Terms})+(\text{Crossed Terms})+(\text{Transient Terms})\nonumber\\
&&-\frac{\Theta(-x_{1})~\Theta(-x_{2})}{\beta_{\phi}}\sum_{j}R_{j}~\frac{r(-z_{j})}{z_{j}}\left[\Theta(x_{1}-x_{2}+t_{1}-t_{2})~\Theta(x_{2}+t_{2}-t_{0})~e^{z_{j}(x_{1}-x_{2}+t_{1}-t_{2})}\right.\nonumber\\
&&\left.+~\Theta(x_{2}-x_{1}+t_{2}-t_{1})~\Theta(x_{1}+t_{1}-t_{0})~e^{-z_{j}(x_{1}-x_{2}+t_{1}-t_{2})}\right].
\end{eqnarray}

At this point, we can exploit one more time the Residue theorem to obtain a final closed form for these terms. It is straightforward to show that first the sum over $j$, as we did for the others sums and taking into account the convergence requirements, can be written as an integral in the complex plane over a curve $\mathcal{C}^{+}=\mathcal{C}_{L}^{+}\bigcup\mathcal{R}^{+}$ where $\mathcal{C}_{L}^{+}$ is a half-infinite circle enclosing the left half of the complex plane and $\mathcal{R}^{+}$ is a straight path over the imaginary axis from the bottom to the top. Therefore, since the integrand function $\frac{r(z)~r(-z)}{(-z^{2})}~e^{z(x_{1}-x_{2}+t_{1}-t_{2})}$ vanishes for $|z|\rightarrow\infty$ with $Re(z)<0$,  the integral over $\mathcal{C}_{L}^{+}$ is null and the integral is directly over the imaginary axis, which can be parametrized as $z=-i\Omega$, finally obtaining:

\begin{eqnarray}
-\sum_{j}R_{j}~\frac{r(-z_{j})}{z_{j}}~e^{z_{j}(x_{1}-x_{2}+t_{1}-t_{2})}&=&\int_{\mathcal{C}^{+}}\frac{dz}{2\pi i}~\frac{r(z)~r(-z)}{(-z^{2})}~e^{z(x_{1}-x_{2}+t_{1}-t_{2})}\nonumber \\
&=&\int_{-\infty}^{+\infty}\frac{d\Omega}{2\pi}~\frac{|r(-i\Omega)|^{2}}{(\Omega^{2})}~e^{-i\Omega(x_{1}-x_{2}+t_{1}-t_{2})},
\end{eqnarray}
where we have used that $r(i\Omega)=r^{*}(-i\Omega)$ for real $\Omega$.

Finally, the proper field contribution reads:

\begin{eqnarray}
\mathcal{A}(x_{1},x_{2},t_{1},t_{2})&+&\mathcal{B}(x_{1},x_{2},t_{1},t_{2})=(\text{Free Field Terms})+(\text{Crossed Terms})+(\text{Transient Terms})\nonumber\\
&&+\frac{\Theta(-x_{1})~\Theta(-x_{2})}{\beta_{\phi}}\int_{-\infty}^{+\infty}\frac{d\Omega}{2\pi}~\frac{|r(-i\Omega)|^{2}}{\Omega^{2}}\left[\Theta(x_{1}-x_{2}+t_{1}-t_{2})~\Theta(x_{2}+t_{2}-t_{0})~e^{-i\Omega(x_{1}-x_{2}+t_{1}-t_{2})}\right.\nonumber\\
&&\left.+~\Theta(x_{2}-x_{1}+t_{2}-t_{1})~\Theta(x_{1}+t_{1}-t_{0})~e^{i\Omega(x_{1}-x_{2}+t_{1}-t_{2})}\right],
\end{eqnarray}
where it is remarkable that it has the form of the long-time contributions considered without demonstration in Refs.\cite{LombiMazziRL,Dorota1992} for the proper field contribution in a steady canonical quantization scheme in the case of the force between two plates, the Casimir-Lifshitz problem.

All in all, we have showed that the long-time limit of a Dirac delta plate of real material have contributions both from the bath as from the field by itself. The result can be extended to other configurations in one dimension ($1+1$) also but the calculations are more complicated. The conclusion seems to be general, i.e., we have showed that a situation including boundaries or, analogously, including vacuum regions, will present not only the contributions from the baths, but also the proper field contribution in the long-time regime. Therefore, this situation shows a new type of scenario, where the long-time regime is steady but it has contributions from two parts of the composite system. Note that the behavior and the steady situation in the case with vacuum regions is radically different from the case of material in the entire space. However, the material contribution, considered separately, behaves in the same way, i.e., the contribution associated to the polarization degrees of freedom is transient and vanish at the steady situation, while the baths' contribution survives and it is part of the long-time regime. The great difference of including boundaries is that it is not the only one that survives. This is due to the fact that while the field tends to vanish inside the material due to dissipation, outside it is fluctuating freely without damping. This makes that the fluctuations outside propagate inside the material and finally reach a steady situation at the long-time regime, when the material has relaxed and the dynamics are reduced to the steady ones, then allowing us to describe the proper contribution effectively by modified vacuum modes as Refs.\cite{LombiMazziRL,Dorota1992} for the Casimir problem.

Therefore, any quantization procedure in the long-time regime, i.e., any quantization scheme at the steady situation, at least for this models in $1+1$, must consider this contribution to obtain the correct results.

Following this train of thought, as a final comment we should note that we have proved this fact in the one dimensional ($1+1$) case, but the conclusion for higher dimensions could change. Although we have not made here the calculation for the $n+1$ case in this scenario including vacuum regions, from the comparison between the reflection and transmission coefficients in Eqs.(\ref{CoefficientsRTnD}) and (\ref{CoefficientsRT1D}), and the Laplace transform of the retarded Green functions in Eqs. (\ref{LaplaceFourierGreenFunctionDirac}) and (\ref{LaplaceFourierGreenFunctionDirac1D}) for both cases, we can note that for the higher dimensions problem we have two branch cuts $\sqrt{z^{2}+k_{\parallel}}$ involving the parallel momentum $k_{\parallel}$ instead of a simple $z$. Therefore, the analytical properties of the Laplace transform are different and the time behavior of the retarded Green function will change critically. It could happen that the proper field contribution vanishes for this higher dimensions case, but the continuity between the actual case of real material and the result obtained in the $n+1$ case for a constant dielectric and arbitrary boundaries in Sec.\ref{CDPL} suggest that the result of this Section is quite general even for the higher dimensions case.

\section{Final remarks}\label{FR}

In this article we have extensively used the CTP approach to calculate a general expression for the time evolution of the expectation value of the energy-momentum tensor components, in a completely general non-equilibrium scenario, for a scalar field in the presence of real materials. The interaction is turned on at an initial time $t_{0}$, coupling the field to the polarization degrees of freedom of a volume element of the material, which is also linearly coupled to thermal baths in each point of the space. Throughout the work, we studied two coupling models between the field and the polarization degrees of freedom. One is the bilinear model, analogous to the one considered in the QBM theory \cite{BreuerPett, CaldeLegg, HuPazZhang}, and the other one is (a more realistic) current-type model, where the polarization degrees of freedom couples to the field's time derivative (as in the EM case interacting with matter).

It is remarkable that the material is free to be inhomogeneous, i.e., its properties (density $\eta_{\mathbf{x}}$, coupling constant to the field $\lambda_{0,\mathbf{x}}$, mass $m_{\mathbf{x}}$, and frecuency $\omega_{\mathbf{x}}$ of the volume elements polarization degrees of freedom) can change with the position. The baths' properties (coupling constant to the polarization degrees of freedom $\lambda_{n,\mathbf{x}}$, mass $m_{n,\mathbf{x}}$, and frecuency $\omega_{n,\mathbf{x}}$ of each bath oscillator) also can change with position, resulting in an effective-position-dependent properties over the volume elements of the material; which are represented by the dissipation and noise kernels $D_{\rm QBM,\mathbf{x}}$ and $N_{\rm QBM,\mathbf{x}}$ after the first integration over the baths' degrees of freedom.

On the other hand, thermodynamical non-equilibrium is included by letting both each volume element and thermal bath to have their own temperature ($\beta_{r_{\mathbf{x}}},\beta_{B,\mathbf{x}}$) by choosing the initial density operators of each part to be thermal states. The field has also its own temperature $\beta_{\phi}$, which analogously comes from the field initial state, but for simplicity in the calculations, we have taken the high temperature approximation, except for the $0+1$ field case, in Sec.\ref{0+1F}, where the calculation can be done for arbitrary temperatures of the field.

It is worth noting that the approach also consider that the material bodies can be of finite extension and have arbitrary shape, i.e., vacuum regions were the field is free are included. All these features are concentrated in the matter distribution function $g(\mathbf{x})$, which takes binary values $1$ or $0$ whether there is material at $\mathbf{x}$ or not.

In the field's high temperature limit, the expectation value of the energy-momentum tensor components are given by Eq.(\ref{TmunuExpValue}), where the Hadamard propagator is defined in Eq.(\ref{HadamardPropagator}). That equation contains the full dynamics of the field correlation, with one contribution clearly associated to the material, which is contained in the third term of Eq.(\ref{HadamardPropagator}); and the other one clearly associated to the field by itself, which is contained in the first two terms.

All in all, the third term is directly associated to the material (polarization degrees of freedom plus thermal baths), represented by the field's noise kernel, which is also separated in two contributions, $\mathcal{N}(x,x')=4\pi\eta_{\mathbf{x}}g(\mathbf{x})~\delta(\mathbf{x}-\mathbf{x}')\left[\mathcal{N}_{r,\mathbf{x}}(\tau,\tau')+\mathcal{N}_{B,\mathbf{x}}(\tau,\tau')\right]$, one associated to the polarization degrees of freedom that define the material and has an effective dissipative (damping) dynamics due to the interaction with the baths (QBM), and other one associated to the baths' fluctuations that, thanks to its (non-damping) dynamics, acts like source of constant temperatures in each point of space and results in an influence over the field, although they are not in direct interaction, but the polarization degrees connect them as a bridge in a non-direct interaction. Both contributions define the field's transient dynamics, while the steady situation seems to be determined only by the baths', which has non-damping dynamics. The relaxation dynamics of the polarization degrees of freedom will make that they will not have any contribution to the long-time regime.

We have shown that the polarization degrees of freedom never contribute to the steady situation, while the bath always do, independently of the situation considered. However, there is a case (Sec.\ref{CDPL}) that both contributions were absent. It is the constant dielectric case, where it is trivially expected that there is no contribution because the material dynamics is suppressed.

On the other hand, the first two terms of Eq.(\ref{HadamardPropagator}) are directly associated to the field initial state. As we have considered the high temperature limit, both terms are linear in the field's initial temperature $\beta_{\phi}$, as is expected. These terms give the transient field evolution from the initial free field over the whole space to the field in interaction with the polarization degrees of freedom of the material in certain regions defined by the distribution function $g(\mathbf{x})$. This evolution involves two dynamical aspects. One is related to the fact that the properties of the boundaries are time-dependent. This adds extra dynamical features associated to the relaxation process of the material, which enter the field dynamics encoded in the field's dissipation kernel $\partial_{tt}^{2\alpha}\mathcal{D}$ for each coupling model that defines the form of the field's retarded Green function $\mathcal{G}_{\rm Ret}$ through Eq.(\ref{EqMotionPhiGeneralized}). In Sec.\ref{CDPL}, this aspect is turned off by suppressing the material's relaxation.

The other dynamical aspect is the adaptation of the free field to the condition of being bounded by the sudden appearance of boundaries. We have shown that this clearly takes place in two scenarios, one was studied in Sec.\ref{FAMB} when the distribution function $g(\mathbf{x})$ has null values for some space point $\mathbf{x}$, i.e., when there are vacuum regions in the particular problem, while the other one is the aforementioned lossless case (Sec.\ref{CDPL}). In both, the aspect is basically the conversion of the field modes from free field modes to interacting (modified) field modes.

This is completely related to the long-time regime of these terms and it is not so easy to analyze. In Secs.\ref{0+1F} and \ref{FIIM}, the proper field contribution vanishes for both coupling models, so there is no interacting (modified) field modes associated to the field's initial state. Tempted to consider that this result applies to all the cases involving material boundaries, in Secs.\ref{CDPL} and \ref{FAMB}, we have shown that the proper field contribution does not vanish in the long-time regime. The result is trivially expected for the constant dielectric permittivity, as there exists steady Casimir force between bodies of constant dielectric permittivity (see Refs.\cite{LombiMazziRL, Dorota1990}) involving a $n+1$ field. Therefore, since this case and the real material including losses and absorption are expected to be continuously connected, as occurs in Sec.\ref{FAMB} for the one dimensional ($1+1$) case, it is suggested that the result found in $1+1$ is quite general and holds for the $n+1$ case with material boundaries (with $n>1$).

Physically, for the case without boundaries, the dissipative (damping) dynamics of the field vanishes at the steady situation as in the QBM case.

For the case with vacuum regions, the field inside the material tends to behave as a damped field, but in the vacuum regions, the field fluctuates without damping. So, there is a competition between the behavior in both regions which will make the field evolves to a modified modes steady field in the long-time regime, beyond the material relaxation, analogously to the case of the constant dielectric properties steady situation, but with frequency-dependent effective properties. The free fluctuations propagate inside the material regions and keep the field in a continuous situation that survives to the long-time regime. In other words, the transient dynamics of these terms will be different from the constant dielectric case. The steady situation will be also different because the frequency dependence of the properties. But, taking into account the considerations made above, the formal expression must be the same as the constant dielectric case by replacing it by the corresponding actual frequency-dependent permittivity (as it happens for the Casimir force for a $1+1$ field in absorbing media \cite{LombiMazziRL}).

Therefore, at least in the $1+1$ case, any quantization scheme of these models at a steady scenario, although involving thermodynamical non-equilibrium, must take into account the proper field contribution besides the expected baths' contribution.

All in all, this is in fact a natural physical conclusion, all the parts of the system having no damping dynamics contribute to the long-time regime. Following this train of thought, is naturally expected that the bath contributes and, since there are regions where the field has no damping dynamics, modified field modes takes place in the long-time proper field contribution.

As future work, for completeness, it should be possible to extend the calculation to the case of arbitrary field's temperature without many complications. More interesting would be investigate the implications of this analysis over the vacuum fluctuations in the real Casimir problem, extending the complete study in $1+1$ scalar field to the $3+1$ EM field. Finally, it would be also interesting to study the heat transfer and other thermodynamical features in situations where the material be thermally inhomogeneous.
\section*{Acknowledgements}
We would like to thank A. J. Roncaglia and F. D. Mazzitelli for useful discussions that stimulated this research.
This work was supported by UBA, CONICET and ANPCyT.

\appendix
\section{The field Wigner functional}\label{A}

The Wigner functional for a quantum field can be defined as in Ref. \cite{MrowMull}:

\begin{eqnarray}
W_{\phi}\left[\phi_{0}(\mathbf{x}),\Pi_{0}(\mathbf{x}),t_{0}\right]=\int\mathcal{D}\varphi(\mathbf{x})~e^{-i\int d\mathbf{x}~\Pi_{0}(\mathbf{x})~\varphi(\mathbf{x})}~\Big\langle\phi_{0}(\mathbf{x})+\frac{1}{2}~\varphi(\mathbf{x})\Big|\widehat{\rho}_{\phi}(t_{0})\Big|\phi_{0}(\mathbf{x})-\frac{1}{2}~\varphi(\mathbf{x})\Big\rangle.
\label{FieldWignerCoordinate}
\end{eqnarray}

It is worth noting that sometimes seems easier to compute the Wigner functional in momentum space. However, it is not so easy. Even though the field $\phi(\mathbf{x})$ is real, it Fourier transform $\phi(\mathbf{p})$ is complex but its real and imaginary parts are not independent, because to have a real field, $\phi(-\mathbf{p})=\phi^{*}(\mathbf{p})$. As in Ref. \cite{MrowMull}, for the Fourier transform, we will treat the real and imaginary parts of $\phi(\mathbf{p})$ as independent variables, but considering $p_{i}\in(0,+\infty)$ for each momentum component instead of $p_{i}\in(-\infty,+\infty)$. This way, the Wigner functional in the momentum space can be defined as:

\begin{eqnarray}
\widetilde{W}_{\phi}\left[\phi_{0}(\mathbf{p}),\Pi_{0}(\mathbf{p}),t_{0}\right]=\int\mathcal{D}\varphi(\mathbf{p})~e^{-i\int_{0}^{+\infty} d\mathbf{p}~\left[\Pi_{0}^{*}(\mathbf{p})~\varphi(\mathbf{p})+\Pi_{0}(\mathbf{p})~\varphi^{*}(\mathbf{p})\right]}~\Big\langle\phi_{0}(\mathbf{p})+\frac{1}{2}~\varphi(\mathbf{p})\Big|\widehat{\rho}_{\phi}(t_{0})\Big|\phi_{0}(\mathbf{p})-\frac{1}{2}~\varphi(\mathbf{p})\Big\rangle,
\label{FieldWignerMomentum}
\end{eqnarray}
with the functional integrations running over real and imaginary components of $\phi(\mathbf{p})$ \cite{MrowMull}. Going from Eq.(\ref{FieldWignerCoordinate}) to (\ref{FieldWignerMomentum}) implies a nontrivial Jacobian $\det\left[\frac{\delta\varphi(\mathbf{x})}{\delta\varphi(\mathbf{p})}\right]$, which does not depends on the fields because the Fourier transformation is a linear mapping, consequently, it appears merely as a new normalization factor of the Wigner functional.

Now, we consider the scalar field initially in thermodynamical equilibrium. Then, the density matrix operator $\widehat{\rho}_{\phi}(t_{0})$ is given by:

\begin{equation}
\widehat{\rho}_{\phi}(t_{0})=\frac{1}{Z}~e^{-\beta_{\phi}\widehat{H}_{0}},
\end{equation}
where $Z$ is the partition function associated to the initial field's hamiltonian $\widehat{H}_{0}$, which can be written as:

\begin{equation}
\widehat{H}_{0}=\int_{0}^{+\infty}d\mathbf{p}\left(\widehat{\Pi}^{\dag}(\mathbf{p})~\widehat{\Pi}(\mathbf{p})+p^{2}~\widehat{\Phi}^{\dag}(\mathbf{p})~\widehat{\Phi}(\mathbf{p})\right).
\end{equation}

In order to check that the hamiltonian is a sum of two harmonic oscillator hamiltonians for each component at fixed $\mathbf{p}$. Thus, taking $\mathbf{p}$ as a label for each pair of oscillators, we can introduce a complete set of energy eigenstates of the two dimensional (isotropic) oscillator $|n_{1},n_{2}\rangle$, writing Eq.(\ref{FieldWignerMomentum}) as:

\begin{eqnarray}
\widetilde{W}_{\phi}\left[\phi_{0}(\mathbf{p}),\Pi_{0}(\mathbf{p}),t_{0}\right]=\sum_{n_{1},n_{2}}\int\mathcal{D}\varphi(\mathbf{p})&&e^{-\int_{0}^{+\infty}d\mathbf{p}\left[i\left(\Pi_{0}^{*}(\mathbf{p})~\varphi(\mathbf{p})+\Pi_{0}(\mathbf{p})~\varphi^{*}(\mathbf{p})\right)+\beta_{\phi}|\mathbf{p}|\right]}\times\nonumber\\
&&\times\Big\langle\phi_{0}(\mathbf{p})+\frac{1}{2}\varphi(\mathbf{p})\Big|n_{1},n_{2}\Big\rangle~\Big\langle n_{2},n_{1}\Big|\phi_{0}(\mathbf{p})-\frac{1}{2}\varphi(\mathbf{p})\Big\rangle.
\end{eqnarray}

The eigenfunctions for the two dimensional (isotropic) harmonic oscillator are given by:

\begin{eqnarray}
\Big\langle\Phi_{R},\Phi_{I}\Big|n_{1},n_{2}\Big\rangle=\left(\frac{\alpha^{2}}{\pi~2^{n_{1}}~n_{1}!~2^{n_{2}}~n_{2}!}\right)^{1/2}H_{n_{1}}\left(\alpha~\Phi_{R}\right)~H_{n_{2}}\left(\alpha~\Phi_{I}\right)~e^{-\frac{\alpha^{2}}{2}\left(\Phi_{R}^{2}+\Phi_{I}^{2}\right)},
\end{eqnarray}
where $\Phi_{R,I}$ is the real or imaginary part of the field, respectively, $H_{n}$ are the Hermite polynomials and $\alpha\equiv\left(\frac{|\mathbf{p}|}{2}\right)^{1/2}$.

Inserting this into the last expression for the Wigner functional in momentum space and using the identity for the Hermite polynomials:

\begin{equation}
\sum_{n}^{\infty}\frac{a^{n}}{n!}~H_{n}(x)~H_{n}(y)=\frac{1}{\sqrt{1-4a^{2}}}~e^{\frac{4axy-4a^{2}(x^{2}+y^{2})}{1-4a^{2}}},
\end{equation}
which holds for $a<1/2$, condition which is satisfied in our case because $a=e^{-\beta_{\phi}|\mathbf{p}|}/2$; one finds that:

\begin{eqnarray}
\widetilde{W}_{\phi}\left[\phi_{0}(\mathbf{p}),\Pi_{0}(\mathbf{p}),t_{0}\right]&=&\int\mathcal{D}\varphi(\mathbf{p})~e^{-i\int_{0}^{+\infty}d\mathbf{p}\left(\Pi_{0}^{*}(\mathbf{p})~\varphi(\mathbf{p})+\Pi_{0}(\mathbf{p})~\varphi^{*}(\mathbf{p})-i\frac{\alpha^{2}}{4}~\varphi^{*}(\mathbf{p})~\varphi(\mathbf{p})-i\alpha^{2}~\phi_{0}^{*}(\mathbf{p})~\phi_{0}(\mathbf{p})-i\beta_{\phi}|\mathbf{p}|\right)}\nonumber\\
&&\times\prod_{\mathbf{p}}\frac{\alpha^{2}}{\pi\left(1-e^{-2\beta_{\phi}|\mathbf{p}|}\right)}~e^{\frac{\alpha^{2}~e^{-\beta_{\phi}|\mathbf{p}|}}{2\left(1-e^{-2\beta_{\phi}|\mathbf{p}|}\right)}\left[4\left(1-e^{-\beta_{\phi}|\mathbf{p}|}\right)~\phi_{0}^{*}(\mathbf{p})~\phi_{0}(\mathbf{p})-\left(1+e^{-\beta_{\phi}|\mathbf{p}|}\right)~\varphi^{*}(\mathbf{p})~\varphi(\mathbf{p})\right]}\nonumber\\
&=&C~e^{\int_{0}^{+\infty}d\mathbf{p}~\alpha^{2}~\tanh\left(\frac{\beta_{\phi}|\mathbf{p}|}{2}\right)\phi_{0}^{*}(\mathbf{p})~\phi_{0}(\mathbf{p})}\int\mathcal{D}\varphi(\mathbf{p})~e^{-i\int_{0}^{+\infty}d\mathbf{p}\left(\Pi_{0}^{*}(\mathbf{p})~\varphi(\mathbf{p})+\Pi_{0}(\mathbf{p})~\varphi^{*}(\mathbf{p})\right)}\nonumber\\
&&\times~e^{\int_{0}^{+\infty}d\mathbf{p}~\frac{\alpha^{2}}{4}~\coth\left(\frac{\beta_{\phi}|\mathbf{p}|}{2}\right)\varphi^{*}(\mathbf{p})~\varphi(\mathbf{p})}.
\end{eqnarray}

where in the coefficient $C$ we have included all the terms which are not functionals of the fields and momenta.

Integrating trivially over the real and imaginary components of $\varphi(\mathbf{p})$, we arrive to the three-dimensional generalization of the Wigner functional in momentum space found in Ref. \cite{MrowMull} for the one-dimensional case:

\begin{eqnarray}
\widetilde{W}_{\phi}\left[\phi_{0}(\mathbf{p}),\Pi_{0}(\mathbf{p}),t_{0}\right]=C~e^{-\frac{\beta_{\phi}}{2}\int d\mathbf{p}~\widetilde{\Delta}_{\beta_{\phi}}(|\mathbf{p}|)\left[\Pi_{0}^{*}(\mathbf{p})~\Pi_{0}(\mathbf{p})+|\mathbf{p}|^{2}~\phi_{0}^{*}(\mathbf{p})~\phi_{0}(\mathbf{p})\right]},
\label{FieldWignerP3DThermal}
\end{eqnarray}
where the thermal weight factor is given by:

\begin{equation}
\widetilde{\Delta}_{\beta_{\phi}}(|\mathbf{p}|)=\frac{2}{\beta_{\phi}|\mathbf{p}|}~\tanh\left(\frac{\beta_{\phi}|\mathbf{p}|}{2}\right).
\label{ThermalWeightMomentum}
\end{equation}
This is an even function of the momentum's absolute value, so in Eq.(\ref{FieldWignerP3DThermal}) the integrals over the momentum components were extended to all the real values.

Finally, knowing the Wigner functional in momentum space, one can easily finds the Wigner functional in coordinate space by writing all the momentum's functions as Fourier transforms of the function in coordinate space. This way, we can write an extension of the result found in \cite{MrowMull}:

\begin{eqnarray}
W_{\phi}\left[\phi_{0}(\mathbf{x}),\Pi_{0}(\mathbf{x}),t_{0}\right]=C'~e^{-\beta\int d\mathbf{x}\int d\mathbf{x}'~\mathcal{H}(\mathbf{x},\mathbf{x}')},
\label{FieldWignerCoordinate}
\end{eqnarray}

\noindent where $C'$ is the normalization constant in coordinate space and the integrand $\mathcal{H}$ is given by

\begin{eqnarray}
\mathcal{H}(\mathbf{x},\mathbf{x}')\equiv\frac{1}{2}~\Delta_{\beta_{\phi}}(\mathbf{x}-\mathbf{x}')\left[\Pi_{0}(\mathbf{x})~\Pi_{0}(\mathbf{x}')+\nabla\phi_{0}(\mathbf{x})\cdot\nabla\phi_{0}(\mathbf{x}')\right],
\end{eqnarray}
where the thermal weight factor in coordinate space is given by:

\begin{eqnarray}
\Delta_{\beta_{\phi}}(\mathbf{x}-\mathbf{x}')=\int\frac{d\mathbf{p}}{(2\pi)^{3}}~e^{-i\mathbf{p}\cdot\left(\mathbf{x}-\mathbf{x}'\right)}~\widetilde{\Delta}_{\beta_{\phi}}\left(|\mathbf{p}|\right).
\end{eqnarray}

It is worth noting that, due to the interchange symmetry of the integrand, the thermal weight factor in coordinate space must be symmetric, i. e., $\Delta_{\beta_{\phi}}(\mathbf{x}'-\mathbf{x})=\Delta_{\beta_{\phi}}(\mathbf{x}-\mathbf{x}')$.

It is remarkable that although the expression of the thermal weight factor in momentum space do not change formally with the number of dimensions we are considering, on the other hand, the thermal factor in coordinate space clearly does.

\end{document}